\documentclass[amsmath, amssymb, preprintnumbers, showpacs, showkeys,aps,prl,superscriptaddress,twocolumn]{revtex4-1}
\usepackage{color} 
\usepackage{hyperref}
\usepackage{amsmath,amssymb,mathrsfs,mathtools}
\usepackage{psfrag}
\usepackage{graphicx}
\usepackage{graphics}
\usepackage{epsfig}
\usepackage{bm}
\usepackage{verbatim,color,ulem}
\usepackage{braket}
\usepackage{verbatim}
\usepackage{dsfont}
\usepackage[table]{xcolor}
\usepackage{ulem}
\usepackage{enumitem}
\newcommand{\br}{\mathbf{r}}

\newcommand{\kr}[1]{\mathbf{k}_0\cdot\mathbf{r}_{#1}}
\newcommand{\Eq}[2]{
	\begin{equation}\label{#1}
	\begin{aligned}
	#2
	\end{aligned} 
	\end{equation}
}
\newcommand{\hs}[2]{\hat\sigma^{#1}_{#2}}	
\newcommand{\Xpct}[1]{\langle#1\rangle}

\newcommand{\ehS}[1]{\Xpct{\hat{S}^{#1}}}
\newcommand{\ehs}[2]{\Xpct{\hat\sigma^{#1}_{#2}}}
\newcommand{\ehss}[4]{\Xpct{\hat\sigma^{#1}_{#2}\hat\sigma^{#3}_{#4}}}
\newcommand{\ehsss}[6]{\Xpct{\hat\sigma^{#1}_{#2}\hat\sigma^{#3}_{#4}\hat\sigma^{#5}_{#6}}}
\newcommand{\ehsr}[2]{\Xpct{\hat{\bar{\sigma}}^{#1}_{#2}}}

\newcommand{\eDt}[2]{e^{i(\Delta_{#1}-\Delta_{#2})t}}

\newcommand{\mg}{\alpha}
\newcommand{\g}{g_\mg}
\newcommand{\me}{\beta}
\newcommand{\e}{e_\me}

\begin{document}
	
\newcommand{\papertitle}{Dipole-dipole frequency shifts in multilevel atoms}
\title{\papertitle}

%###################################################################%
%###################          AUTHORS       ########################%
%###################################################################%
	
	\author{A. Cidrim}
	\affiliation{Departamento de F\'isica, Universidade Federal de S\~ao Carlos, 13565-905 S\~ao Carlos, S\~ao Paulo, Brazil}
	\affiliation{JILA, NIST, Department of Physics, University of Colorado, Boulder, CO 80309, USA}
	\affiliation{Center for Theory of Quantum Matter, University of Colorado, Boulder, CO 80309, USA}
	\author{A. Pi\~{n}eiro Orioli}
	\affiliation{JILA, NIST, Department of Physics, University of Colorado, Boulder, CO 80309, USA}
	\affiliation{Center for Theory of Quantum Matter, University of Colorado, Boulder, CO 80309, USA}
	\author{C. Sanner} 
	\affiliation{JILA, NIST, Department of Physics, University of Colorado, Boulder, CO 80309, USA}
	\author{R. B. Hutson}
	\affiliation{JILA, NIST, Department of Physics, University of Colorado, Boulder, CO 80309, USA}
	\author{J. Ye}
	\affiliation{JILA, NIST, Department of Physics, University of Colorado, Boulder, CO 80309, USA}
	\author{R. Bachelard}
	\affiliation{Departamento de F\'isica, Universidade Federal de S\~ao Carlos, 13565-905 S\~ao Carlos, S\~ao Paulo, Brazil}
	\author{A. M. Rey}
	\affiliation{JILA, NIST, Department of Physics, University of Colorado, Boulder, CO 80309, USA}
	\affiliation{Center for Theory of Quantum Matter, University of Colorado, Boulder, CO 80309, USA}

%###################################################################%
%##################          ABSTRACT       ########################%
%###################################################################%	
	
\begin{abstract}
	
Dipole-dipole interactions lead to frequency shifts that are expected to limit the performance of next-generation atomic clocks. In this work, we compute dipolar frequency shifts accounting for the intrinsic  atomic multilevel structure in standard Ramsey spectroscopy. When interrogating the transitions featuring the smallest Clebsch-Gordan coefficients, we find that a simplified two-level treatment becomes inappropriate, even in the presence of large Zeeman shifts. For these cases, we show a net suppression of dipolar frequency shifts and the emergence of dominant non-classical effects for experimentally relevant parameters. Our findings are pertinent to current generations of optical lattice and optical tweezer clocks, opening a way to further increase their current accuracy, and thus their potential to probe fundamental and many-body physics.
\end{abstract}
	
\maketitle
	
%###################################################################%
%################          INTRODUCTION       ######################%
%###################################################################%

\textit{Introduction.---}Current optical atomic clocks have reached unprecedented precision and accuracy~\cite{Nicholson2015,Ludlow2015,Schioppo2016,Brown2017,Campbell2017,McGrew2018,Oelker2019,Norcia2019,Young2020,Lange2021}, making them cutting-edge platforms for many technological applications and for the exploration of many-body~\cite{Rey2014,Zhang2014,Scazza2014,Livi2016,Hofrichter2016,KanaszNagy2018,Goban2018,Sonderhouse2020,Schafer2020,Heinz2020} and fundamental physics~\cite{Chou2010,Grotti2018,Sanner2019,Derevianko2014,Kennedy2020}. The reduction of noise in atomic detection and laser stabilization in such systems has allowed measurements of the atomic transition with submillihertz resolution~\cite{McGrew2018,Bothwell2019,Marti2018}. At this point, dipole-dipole interactions between the atoms are expected to play an important role, in the form of induced density dependent shifts in the measured atomic transition frequency. Simple two-level models have been applied to quantitatively determine these dipolar shifts~\cite{Chang2004,Ostermann2012,Ostermann2013,Kramer2016,Henriet2019,Qu2019,Liu2020}, but, in reality, atoms have a complex internal multilevel structure which has to be taken into account. This calls for a deeper understanding of the role of multiple internal levels in dipolar systems~\cite{Hebenstreit2017,Munro2018,Asenjo2019,Orioli2019,Orioli2020}, which is also relevant for applications in quantum simulators~\cite{Gorshkov2010,Rey2014,Goban2018,Sonderhouse2020} and quantum computing~\cite{Daley2011,Kiktenko2015,Godfrin2018}. 

In this work, we investigate dipolar frequency shifts experienced by arrays of multilevel atoms in a Ramsey spectroscopy protocol. In general, the strength of dipolar interactions is set by the magnitude of the transition's dipole moment, which is proportional to a Clebsch-Gordan coefficient (CGC). However, in multilevel atoms the dependence of the dipolar shift on the choice of transition is more complex. This is because the CGC between two specific states not only sets the strength of the dipole couplings, but also affects the coupling strength to nearby levels. Specifically, transitions with low (high) CGC feature  a stronger (weaker) decay to and interactions with their neighbouring states.

Our results show that the magnitude of the dipolar frequency shift is mainly controlled by the CGC of the {\it interrogated} levels. Therefore, one can strongly suppress dipolar shifts by selectively choosing the levels with the smallest CGC. We also find that interactions with nearby levels can significantly modify the shift. Specifically, we show that a full multilevel calculation is necessary when the CGC of the interrogated transition is small, whereas simplified two-level models are accurate when the CGC is large. Surprisingly, the relevance of the multilevel structure holds even in the presence of strong magnetic fields, under which the large Zeeman shifts suppress exchange with nearby levels.
Moreover, we find that the suppression of the shift from small CGC leads to an increased relative importance of beyond-mean-field effects for specific experimentally relevant array geometries and laser wave vector  configurations. In short, our work offers a simple way for current experiments to reduce dipolar shifts by almost two orders of magnitude, while at the same time drawing theorists' attention to the important yet largely neglected role of internal levels in many-body dipolar systems.
	
%###################################################################%
%###########		MULTILEVEL COUPLED DIPOLE MODEL		############%
%###################################################################%

\begin{figure}[t]
	\centering
	\includegraphics[width=\columnwidth]{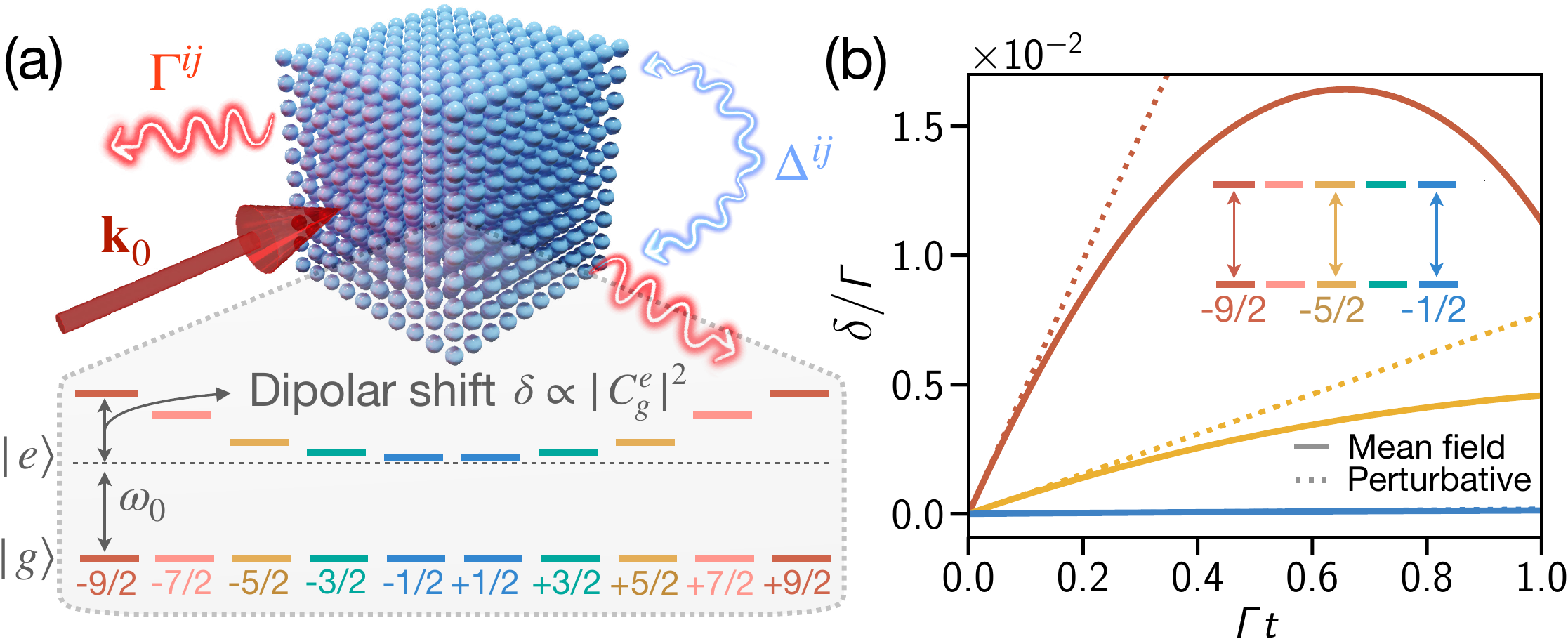} 
	\caption{(a) Ramsey spectroscopy for multilevel atoms with internal level structure $F_g=F_e=9/2$ in an optical lattice. The atoms are prepared in a superposition of a particular ground and excited pair of states $\g$ and $\e$ by a laser with wave vector $\mathbf{k}_0$ and pulse area $\theta$. During a dark time $t$, the atoms interact via coherent and incoherent dipole-dipole processes, $\Delta^{ij}$ and $\Gamma^{ij}$, respectively. This induces a frequency shift $\delta^{\g\e}\sim|C^{\e}_{\g}|^2$ controlled by the CGC of the interrogated transition. The schematic form of the dipolar shift corresponding to interrogating $\pi$-polarized transitions (colored according to their CGC) is depicted here . (b) Shift from the Ramsey protocol addressing three different $\pi$-polarized transitions for a 3D lattice of spacing $d=7\lambda/12$ with $N=10^3$ atoms and $\theta=\pi/2$. The dipolar shift is highly suppressed for the transition with smallest CGC.} \label{Fig:scheme}
\end{figure}

\textit{Multilevel coupled dipole model.---}We consider a system of $N$ point-like atoms pinned in a deep optical lattice or a tweezer array with unity occupation, always in their motional ground state. We assume that each atom $i$ has a multilevel internal structure of ground and excited manifolds, $g$ and $e$, with respective total angular momenta $F_g$ and $F_e$. There are thus $(2F_a+1)$ hyperfine states $|a_m\rangle_i\equiv|a,F_a,m\rangle_i$ with angular momentum projections $m\in\left[-F_a,F_a\right]$, for each manifold $a\in \{g,e\}$. The photon-mediated interaction between the atoms occurs via both coherent exchange and incoherent decay of excitations [see Fig.~\ref{Fig:scheme}(a)], and the dipole dynamics can be modelled by a \textit{multilevel coupled dipole master equation}~\cite{Lehmberg1970,Gross1982,James1993,Orioli2019,Orioli2020} $\dot{\hat\rho}=-i\left[\hat{H},\hat\rho(t)\right]+\mathcal{L}(\hat\rho)$ ($\hbar=1$), where
\begin{align}
\label{Eq:coherent_part}
\hat{H}=&-\sum_{i, j}\Delta_{g_me_n,g_{m'}e_{n'}}^{ij}\hs{e_{n}g_{m}}{i}\hs{g_{m'}e_{n'}}{j},\\
\label{Eq:incoherent_part}
\mathcal{L}(\hat{\rho})=&\,\sum_{i,j}\Gamma_{g_me_n,g_{m'}e_{n'}}^{ij}\Big(2\,\hs{g_{m'}e_{n'}}{j}\hat{\rho}\,\hs{e_ng_m}{i}\nonumber\\[-5pt]
&\,\qquad\qquad\qquad-\left\{\hs{e_ng_m}{i}\hs{g_{m'}e_{n'}}{j},\hat{\rho}\right\}\Big),
\end{align}

\noindent and $\hat{\sigma}_i^{a_mb_n}=|a_m\rangle_i\langle b_n|_i$. For a two-level atom, these operators become the usual raising/lowering Pauli operators. For clarity, we have used Einstein notation for levels in the equations above (i.e., repeated indices $m,m',n$, or $n'$ are summed). The terms proportional to $\Delta^{ij}_{g_me_n,g_{m'}e_{n'}}$ and $\Gamma^{ij}_{g_me_n,g_{m'}e_{n'}}$ characterize the elastic and dissipative components of the dipolar interactions  and their amplitudes relate to the free-space electromagnetic Green's tensor $\mathbf{G}_{ij}\equiv \mathbf{G}(\br_i-\br_j)$ of an oscillating point dipole at position $\br_j$ according to
\Eq{Eq:Deltas_Gammas}{\Delta^{ij}_{g_me_n,g_{m'}e_{n'}}&\equiv C_{g_m}^{e_n}\mathbf{e}^{*}_{n-m}\cdot \mathrm{Re}\left\{\mathbf{G}_{ij}\right\}\cdot C_{g_{m'}}^{e_{n'}}\mathbf{e}_{n'-m'}, \\	
\Gamma^{ij}_{g_me_n,g_{m'}e_{n'}}&\equiv C_{g_m}^{e_n}\mathbf{e}^{*}_{n-m}\cdot \mathrm{Im}\left\{\mathbf{G}_{ij}\right\}\cdot C_{g_{m'}}^{e_{n'}}\mathbf{e}_{n'-m'},}	
where $C_{g_m}^{e_n}\equiv \langle F_g, m; 1, n - m | F_e, n \rangle$ is the CGC of the transition $g_m\leftrightarrow e_n$ with polarization vector $\mathbf{e}_{n-m}$. We define the spherical basis $\mathbf{e}_{0}=\hat z$, $\mathbf{e}_{\pm1}=\mp(\hat x \pm i \hat y)/\sqrt{2}$. The vacuum Green's tensor is given by $\mathbf{G}(\br)=(3\Gamma/4)(e^{ik_0r}/(k_0r)^3)\Big[\left(k_0^2r^2+ik_0r-1\right)\mathds{1}-\left(k_0^2r^2+i3k_0r-3\right)\hat{\br}\otimes\hat{\br}\Big]$, where $\hat{\br}=\br/r$, $r=|\mathbf{r}|$. $\Gamma= \left|d_{eg}\right|^2 k_0^3/\left[3\pi\hbar\epsilon_0 (2F_e+1)\right]$ is the total spontaneous decay rate, $d_{eg}$ the radial dipole matrix element, $k_0=\omega_0/c=2\pi/\lambda$ the atomic transition wavenumber and $\epsilon_0$ the vacuum permittivity. For $i=j$, the coherent interaction coefficient is $\Delta^{ii}_{g_me_n,g_{m'}e_{n'}}=0$ and the incoherent term reduces to the single-particle spontaneous decay term $\Gamma^{ii}_{g_me_n,g_{m'}e_{n'}}=\delta_{n-m,n'-m'}C_{g_m}^{e_n}C_{g_{m'}}^{e_{n'}}\Gamma/2$. Note that the total decay rate $\Gamma_{e_n}\equiv 2\sum_m\Gamma^{ii}_{g_me_n,g_me_n}=\Gamma$ is the same for any excited state $e_n$ because of the sum rule $\sum_m |C_{g_m}^{e_n}|^2=1$.

%###################################################################%
%#######	RAMSEY SPECTROSCOPY WITH MULTILEVEL ATOMS     ##########%
%###################################################################%

\textit{Ramsey spectroscopy with multilevel atoms.---}We investigate the effect of the atomic multilevel nature on the following Ramsey spectroscopy protocol assuming, at first, zero external magnetic field. We start by selecting a pair of states $\g$ and $\e$, driving the transition between them with a resonant laser of pulse area $\theta$, wave vector $\mathbf{k}_0$, and polarization $\boldsymbol{\epsilon}$. The laser drive is assumed to be much stronger than the interaction energies, such that it creates an uncorrelated coherent superposition $\ket{\Psi_{\g,\e}} = \bigotimes_{j} \Big(\cos (\theta/2) \ket{\g}_j + e^{i\mathbf{k}_0\cdot \br_j}\sin (\theta/2) \ket{\e}_j\Big)$. We hereafter consider $\theta=\pi/2$, as generally used in clock experiments, or $\theta=\pi/4$, as the latter can lead to more pronounced and thus easily observable dipolar shifts. Then, the system evolves freely for a dark time $t$. By analogy with two-level systems, we define $\ehS{y}\equiv \mathrm{Im}\{\ehS{\e\g }\}$ and $\ehS{x}\equiv \mathrm{Re}\{\ehS{\e\g }\}$, where the multilevel collective spin operator (under the appropriate gauge transformation that removes the phase $\kr{j}$ imprinted by the laser on atom $j$) reads $\hat{S}^{\e\g}=\sum_j e^{i\mathbf{k}_0\cdot \br_j}\hat{\sigma}^{\e\g }_{j}$. The collective vector precesses around the $z$-direction of the Bloch sphere and accumulates an azimuthal phase as a result of the dipole-dipole interactions. The corresponding time-dependent frequency shift is defined as \Eq{Eq:shift}{\delta^{\g\e}(t) \equiv \frac{1}{2\pi t} \arctan\frac{\ehS{y}(t)}{\ehS{x}(t)}.} 
Dipolar interactions also lead to a reduction of the contrast $\mathcal{C}^{\g\e}(t)\equiv \frac{1}{N} \sqrt{\ehS{x}^2(t)+\ehS{y}^2(t)}$.

We employ three different types of approximations to investigate this multilevel many-body system:
\begin{enumerate}[leftmargin =0cm, listparindent=2cm, parsep=0.03cm, topsep=0.08cm]
    \item[] i) a short-time perturbative expansion, valid for $t\ll \Gamma^{-1}$, where operators are expanded as $\langle\hat{\mathcal{O}}\rangle \approx \langle\hat{\mathcal{O}}\rangle_0 + \langle\hat{\mathcal{O}}\rangle_1 t + \langle\hat{\mathcal{O}}\rangle_2 t^2/2$, allowing us to compute 
the dipolar frequency shift at first order in time, i.e., $\delta^{\g\e}(t)\approx \delta^{\g\e}_0 + \delta^{\g\e}_1 t$;
\item[] ii) a mean-field (MF) approximation,
which neglects the build up of quantum correlations by approximating two-atom correlators as $\langle\hat{\sigma}^{ab}_i\hat{\sigma}^{cd}_j\rangle\approx \langle\hat{\sigma}^{ab}_i\rangle\langle\hat{\sigma}^{cd}_j\rangle$ (for $i\neq j$);
\item[] iii) a second-order cumulant expansion, which factorizes three-point (and higher-order) correlations in terms of one- and two-point functions $\langle\hat{\sigma}^{ab}_{i}\hat{\sigma}^{cd}_{j}\hat{\sigma}^{ef}_{k}\rangle\approx-2\langle\hat{\sigma}^{ab}_{i}\rangle\langle\hat{\sigma}^{cd}_{j}\rangle\langle\hat{\sigma}^{ef}_{k}\rangle +  \langle\hat{\sigma}^{ab}_{i}\rangle\langle\hat{\sigma}^{cd}_{j}\hat{\sigma}^{ef}_{k}\rangle+\langle\hat{\sigma}^{cd}_{j}\rangle\langle\hat{\sigma}^{ab}_{i}\hat{\sigma}^{ef}_{k}\rangle+ \langle\hat{\sigma}^{ef}_{k}\rangle\langle\hat{\sigma}^{ab}_{i}\hat{\sigma}^{cd}_{j}\rangle$ (with $i$, $j$, $k$ all different).
\end{enumerate}
Due to the large number of equations to solve, in the MF and cumulant calculations we further assume that, when addressing a transition $\g\leftrightarrow\e$, only $\g$, $\e$, and their adjacent levels (i.e.,~$g_{\alpha\pm 1}$ and $e_{\beta\pm 1}$) play a relevant role in the dynamics. We have checked on smaller systems that the neglected levels have no significant effect on the frequency shift over the dark times considered ($\Gamma t< 1$)~\cite{SupMat}.

%###################################################################%
%########## 	SHORT-TIME PERTURBATIVE EXPANSION     ##############%
%###################################################################%

\textit{Short-time perturbative expansion.---}To gain physical intuition of the problem, we analytically derive short-time expressions for the shift. The zero-order shift reads
\Eq{Eq:zero_order_shift}{\delta_0^{\g\e}& = -\frac{\cos\theta}{2\pi N}\sum\limits_{i,j\neq i}U^{ji}_{\g\e},}
where we have defined $U^{ji}_{\g\e}\equiv \Gamma^{ji}_{\g\e,\g\e}\sin(\kr{ij})+ \Delta^{ji}_{\g\e,\g\e}\cos(\kr{ij})$. Physically, the term $U^{ji}_{\g\e}$ describes the classical interaction energy between two oscillating dipoles at positions $\br_{i}$ and $\br_{j}$~\cite{Chang2004}, where both coherent and incoherent processes contribute. At this order, only the transition between $\g$ and $\e$, directly driven by the pulse, is involved  and the MF treatment is exact. Furthermore, $\delta_0^{\g\e}$ is proportional to $|C_{\g}^{\e}|^2$ [see Eq.~(\ref{Eq:Deltas_Gammas})], so that the multilevel system differs from two-level atoms~\cite{Chang2004} via a renormalization by the CGC. Note that the zero-order shift vanishes for a $\theta=\pi/2$ pulse, as the dipole-dipole induced precession of the collective Bloch vector requires a non-zero $\langle\hat{S}^z\rangle$ component. 

The next-order correction does involve other levels and is given by
\Eq{Eq:first_order_shift}{\delta^{\g\e}_1
&=-\frac{1}{2\pi N}\sum_{i,j\neq i}\Bigg\{U^{ji}_{\g\e}\widetilde{\Gamma}_{\g\e}\left(\theta\right)
\\+&\sum_{p}\Bigg(\sum_{k\neq i,j}W^{kji}_p\left(\theta\right)+\sum_{p'}Q^{ji}_{p,p'}\left(\theta\right)\Bigg)\Bigg\},
}
with $p$ and $p'$ referring to polarizations~\cite{SupMat}.

The first contribution in Eq.~(\ref{Eq:first_order_shift}) is similar to the zero-order shift. The $\cos\theta$, however, is replaced by $\widetilde{\Gamma}_{\g\e}\left(\theta\right)$, which contains a collective contribution and an explicit dependence on the CGC of the transition interrogated. The $W^{kji}_p(\theta)$ are two-photon coherent and incoherent processes between three different atoms, where one of the contributing transitions is always $\g\leftrightarrow\e$. Thus, these terms are proportional to at least $|C_{\g}^{\e}|^2$. The $Q^{ji}_{p,p'}(\theta)$ terms correspond to processes involving two atoms only, yet not necessarily from the $\g\leftrightarrow\e$ transition. As two-photon processes, they nevertheless contain the product of four CGCs and, as we shall discuss later, they carry beyond-mean-field contributions.

%###################################################################%
%########## 	SUPPRESSION OF THE FREQUENCY SHIFT     #############%
%###################################################################%

\textit{Suppression of the frequency shift.---}Although our conclusions are valid for generic multilevel systems, in  this work we focus our analysis on the case of $^{87}$Sr, given its metrological relevance for atomic clocks~\cite{Rey2014,Nicholson2015,Campbell2017,Goban2018}. More specifically, we assume multilevel atoms with $F_g=F_e=9/2$, organized in a 2D or 3D array with magic-wavelength spacing $d=7\lambda/12$~\cite{Okaba2014}, see Fig.~\ref{Fig:scheme}(a). For simplicity, we will hereafter consider addressing $\pi$-polarized transitions (i.e., $\alpha=\beta$), where the quantization axis is defined by the laser polarization $\boldsymbol\epsilon$. For this system it is important to know that the CGC for $\pi$-polarized transitions scales as $C^{e_m}_{g_m}\propto m$, i.e.,~it is largest for $\pm9/2$ and smallest for $\pm1/2$.

A direct consequence of the zero- and first-order terms' dependence on the CGC is that the shift can be strongly suppressed by choosing the appropriate transition, i.e., the one with the lowest CGC. This effect is illustrated in Fig.~\ref{Fig:scheme}(b), where the dark-time evolution of the shift in a 3D array is monitored for three different transitions, $\alpha=-9/2,-5/2$, and $-1/2$. As a consequence of the scaling with the CGC, the shift is reduced by a factor $81$ for the $\alpha=-1/2$ transition, as compared to $\alpha=-9/2$. Note that the suppression remains valid even at longer times beyond the regime of validity of the short-time expansion. The decay of the contrast $\mathcal{C}^{\g\e}(t)$ also shows a scaling with the CGC, which leads to suppressed sub/superradiance effects for $\alpha=-1/2$ \cite{SupMat}. Furthermore, the excellent agreement in Fig.~\ref{Fig:scheme}(b) between the short-time expressions (dotted lines) and the MF dynamics (full lines) until $\Gamma t\approx 0.2$ shows that, on these time scales, the beyond-MF terms [$Q$ in Eq.~(\ref{Eq:first_order_shift})] do not contribute substantially.

 \begin{figure}[t!]
	\centering
	\includegraphics[width=\columnwidth]{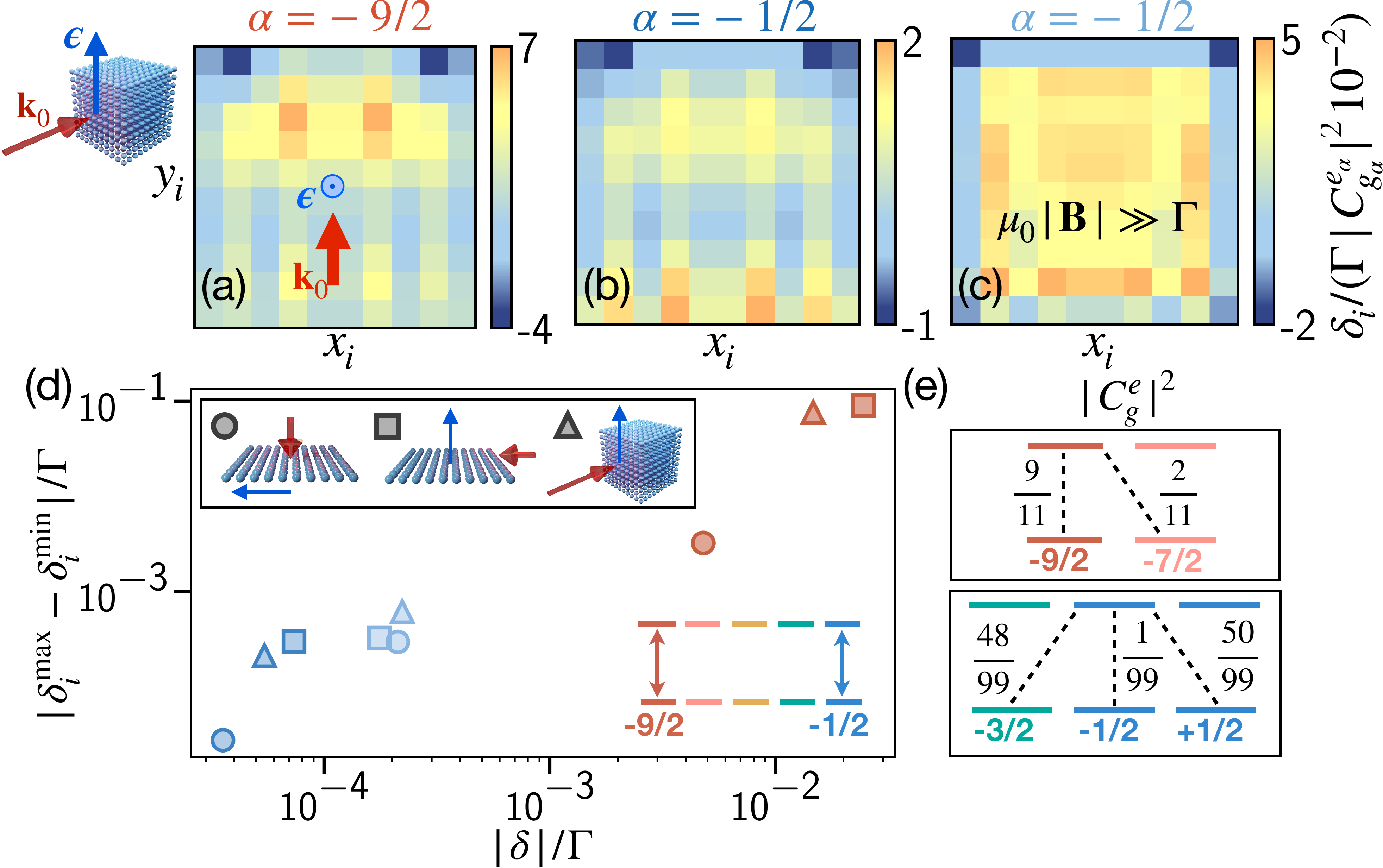} 
    \caption{Local frequency shifts $\delta_i$ for a 3D lattice with $N=10^3$ atoms, interrogated by a laser (as shown in the scheme to the left of the first panel) with a pulse area of $\pi/2$ and a dark time of $\Gamma t = 0.3$ for transitions (a) $g_{-9/2}\leftrightarrow e_{-9/2}$, (b) $g_{-1/2}\leftrightarrow e_{-1/2}$, and (c) $g_{-1/2}\leftrightarrow e_{-1/2}$ in the presence of a large magnetic field $\mathbf{B}$. The shift is calculated using Eq.~(\ref{Eq:first_order_shift}) and is averaged along $\boldsymbol{\epsilon}$, with the resulting contribution at positions $(x_i,y_i)$. Shifts are rescaled by the corresponding CGC squared and by an overall $10^{-2}$ factor. (d) Absolute value of the difference between maximum and minimum of the local shift ($\delta_i^{\text{max}}$ and $\delta_i^{\text{min}}$) versus magnitude of the global shift $|\delta|$ for different geometries with $N\sim 10^3$. Symbols represent configurations shown in the inset. The light-blue symbols correspond to the large-$|\mathbf{B}|$ limit for $\alpha=-1/2$. (e) CGC squared for different $\sigma^\pm$ and $\pi$-transitions. 
	}  
	\label{Fig:local_shifts}
\end{figure}

Further insight is provided by the {\it local} dipolar shift patterns $\delta^{\g\e}_i \equiv \frac{1}{2\pi t} \arctan(\Xpct{\hat s^{y}_i}/\Xpct{\hat s^{x}_i})$ [single-particle counterpart of Eq.~(\ref{Eq:shift})]. Local density shifts directly encode the anisotropic and geometry-dependent character of dipolar interactions, so they can provide further information on the importance of the multilevel structure in  experimentally relevant array geometries. Moreover, local density shifts are amenable for experimental observation via imaging spectroscopy~\cite{Marti2018}, since they are insensitive to laser drifts which are common for all atoms in the array. In Fig.~\ref{Fig:local_shifts}(a-c), we present the local shifts obtained with $\pi/2$ pulses on 3D lattices with $N=10^ 3$ for $\alpha=-9/2$ and $-1/2$.

The magnitude of the shifts shows again an overall suppression with $|C^{e_\alpha}_{\g}|^2$, which we emphasize by rescaling the plots as $\delta_i/(\Gamma|C^{e_\alpha}_{\g}|^2 10^{-2})$. However, the dipolar patterns of the $-9/2$ and $-1/2$ transitions feature distinguishable spatial profiles: the different dispositions of maxima and minima of the local shifts go beyond the mere $|C^{e_\alpha}_{\g}|^2$ scaling.

The local shifts of the $-1/2$ transition reveal a more pronounced sensitivity to the multilevel structure compared to the $-9/2$ case, as confirmed by simulations of pure two-level atoms, which show patterns that are almost indistinguishable from the $-9/2$ case~\cite{SupMat}. This is because the $-1/2$ $\pi$-transition has a small CGC compared to the adjacent $\sigma^\pm$-transitions, whereas for the $-9/2$ $\pi$-transition the opposite is true, see~Fig.~\ref{Fig:local_shifts}(e). Therefore, nearby levels play a more important role in the $-1/2$ case. 

\begin{figure}[!t]
	\centering
	\includegraphics[width=\columnwidth]{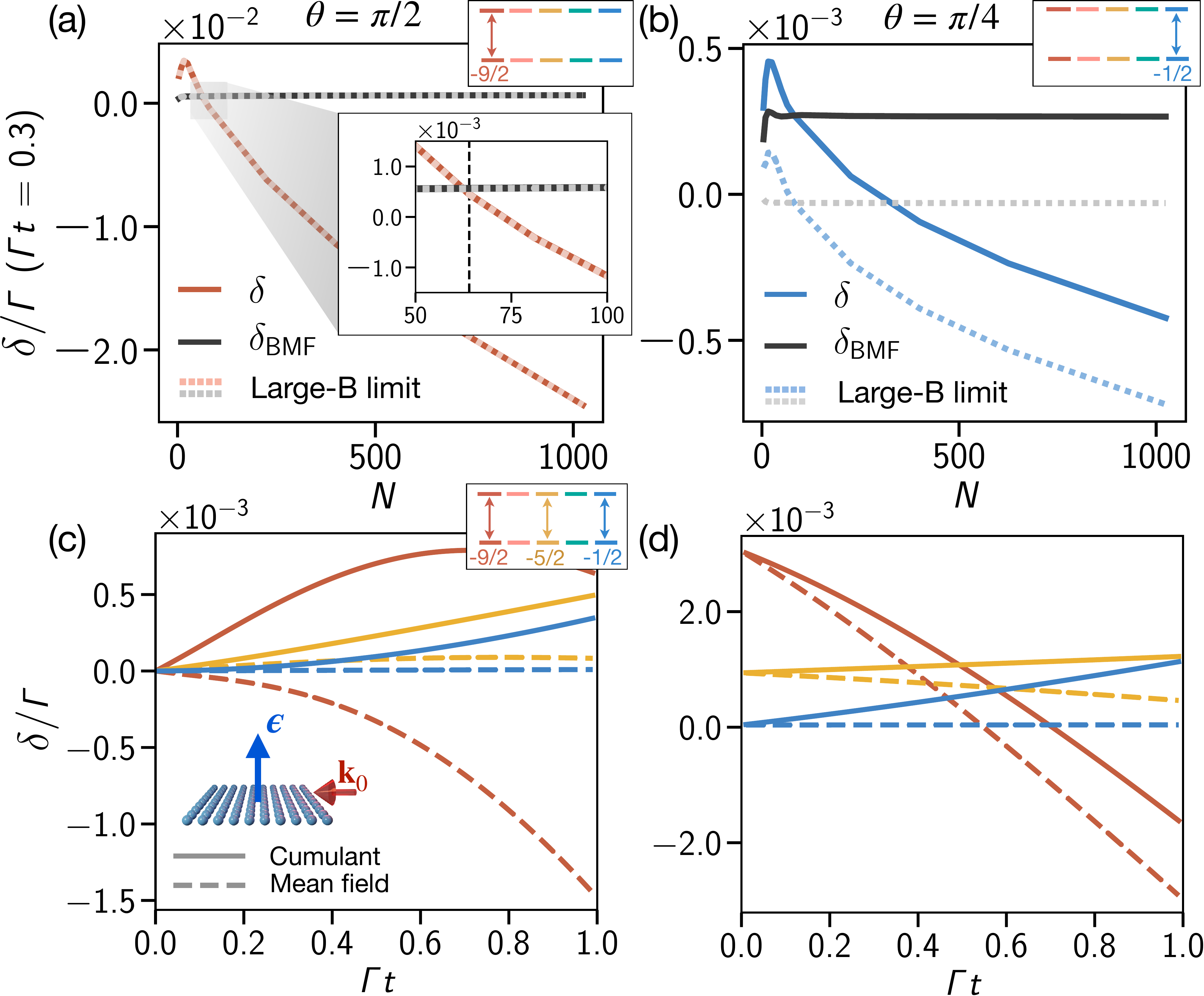} 
	\caption{Global shift for 2D arrays of atoms with laser configuration shown in (c). (a,b) $N$-scaling of the total shift from the short-time expansion (red/blue) and beyond-MF contribution (black) at $\Gamma t=0.3$: (a) shows the $g_{-9/2}\leftrightarrow e_{-9/2}$ transition with $\pi/2$, and (b) the $g_{-1/2}\leftrightarrow e_{-1/2}$ transition with $\pi/4$. The dotted, light-colored lines correspond to the large-magnetic-field limit. The inset of (a) shows the zoomed-in region where beyond-MF corrections become comparable to the total shift. (c,d) Dipolar shift $\delta$ as a function of the dark time: cumulant (full lines) against MF (dashed lines) approximations for the $g_{-9/2}\leftrightarrow e_{-9/2}$ (red), $g_{-5/2}\leftrightarrow e_{-5/2}$ (yellow), and $g_{-1/2}\leftrightarrow e_{-1/2}$ (blue) transitions. Simulations performed for $N=8^2$ atoms and using (c) a $\pi/2$ and (d) $\pi/4$ pulse.}
	\label{Fig:shift_BMF_scaling} 
\end{figure}

The local patterns also sensitively depend on the geometry and laser wave vector. Figure~\ref{Fig:local_shifts}(d) reveals that an appropriate choice of the geometry and laser wave vector allows one to further reduce the shift. Related works have verified this geometry-dependent suppression for two-level systems~\cite{Chang2004,Kramer2016,Liu2020}. Our main finding in this regard is that the dipolar shift saturates for large $N$ in all cases shown, except in 2D when the laser wave vector is parallel to the atomic plane, see Figs.~\ref{Fig:shift_BMF_scaling}(a) and (b). This is because in the latter configuration, all the dipoles align perpendicular to the plane and the corresponding dipolar interactions depend only on the distance between atoms, and not on their orientation~\cite{SupMat}.

%###################################################################%
%###########      CORRECTIONS FROM MAGNETIC FIELDS     #############%
%###################################################################%

\textit{Role of magnetic fields.---}Optical clock experiments are typically conducted under a bias magnetic field $\mathbf{B}$ (along the quantization axis) that allows to spectroscopically address specific transitions. This leads to a Zeeman shift of order $\mu_0|\mathbf{B}|$ (with $\mu_0\equiv\mu_B/\hbar$ and $\mu_B$ the Bohr magneton) for the $\g\leftrightarrow e_\alpha$ transition considered, which trivially adds to the zero-order expression of Eq.~(\ref{Eq:zero_order_shift}) and can be removed in the appropriate rotating frame. However, magnetic fields can non-trivially affect dipolar shifts at higher orders.

If the magnetic field is weak (i.e., $\mu_0|\mathbf{B}|\lesssim \Gamma$), we find the above results on the dipolar shift are only weakly affected at late times. This is because the first-order correction, Eq.~(\ref{Eq:first_order_shift}), turns out to be independent of the magnetic field~\cite{SupMat}. In contrast, strong magnetic fields ($\mu_0|\mathbf{B}|\gg \Gamma$) can significantly alter the short-time behavior of the shift. Large Zeeman shifts effectively suppress exchange interactions involving off-resonant transitions. In other words, $\Delta^{ij}_{g_me_n,g_{m'}e_{n'}}=\Gamma^{ij}_{g_me_n,g_{m'}e_{n'}}=0$ unless $m=m'$ and $n=n'$ (assuming different $g$-factors for the ground and excited manifolds). This leads to an effective 4-level (or 3-level) system composed of $e_\alpha$, $\g$, and the ground levels adjacent to it.
In this limit, almost all terms in the first-order expression, Eq.~(\ref{Eq:first_order_shift}), involving transitions different from $\g\leftrightarrow e_\alpha$ are suppressed, except for terms with $p=p'$ appearing in $Q^{ji}_{p,p'}$~\cite{SupMat}.

Consistently with the discussion above, we find that the modification of the shift strongly depends on the CGC of the addressed transition. For $-9/2$ neither the global nor the local shifts are substantially altered~\cite{SupMat} [see, e.g.,~Fig.~\ref{Fig:shift_BMF_scaling}(a)]. In contrast, for $-1/2$ both the local shift pattern [cf.~Figs.~\ref{Fig:local_shifts}(b) and (c)] as well as the global shift [Fig.~\ref{Fig:shift_BMF_scaling}(b)] are significantly modified under a large $|\mathbf{B}|$. Despite this, the global shift remains suppressed by the small CGC as found for small $|\mathbf{B}|$.

%###################################################################%
%##############      BEYOND-MEAN-FIELD EFFECTS     #################%
%###################################################################%

\textit{Beyond-mean-field effects.---}An important consequence of the strong shift suppression is that higher-order, non-classical terms can have a contribution comparable to the lowest-order, semi-classical ones. The zero-order shift, Eq.~\eqref{Eq:zero_order_shift}, is perfectly described by the MF approach, yet the $Q$ terms of the first-order, Eq.~\eqref{Eq:first_order_shift}, are not. More specifically, the difference between the shift given by the exact, first-order perturbative equations and the MF approximation reads
\Eq{Eq:bmf}{\delta^{\g\e}_{\text{BMF}}\equiv \frac{1}{2\pi N}\sum_{i,j\neq i}\sum_{p}\Bigg\{\sum_{p'}Q^{ji}_{p,p'}\left(\theta\right)-W_{\text{MF},p}^{ji}\left(\theta\right)\Bigg\},
}
where $W_{\text{MF},p}^{ji}\left(\theta\right)$ is a MF-only term related to $W^{kji}$ from Eq.~(\ref{Eq:first_order_shift})~\cite{SupMat}. In general, we find beyond-MF effects to be relevant in cases (but not in {\it every} case) where either the system is small or when a transition with small CGC is addressed. Therefore beyond-MF effects could be relevant for recent tweezer clocks experiments which operate with relative small systems and enjoy almost a minute-long coherence time~\cite{Norcia2019,Young2020}.

Figure~\ref{Fig:shift_BMF_scaling}(c) shows the effect of beyond-MF terms for a small $8^2$ lattice, driven by a $\pi/2$ pulse with a polarization orthogonal to it. There, the dynamics predicted by MF (dashed lines) substantially deviates from the cumulant result (solid lines) for all transitions considered. The relevance of the beyond-MF term in these cases lies in the comparably small magnitude of the MF part. Figure~\ref{Fig:shift_BMF_scaling}(a) shows that for this system size the total shift happens to be close to zero. On the contrary, at large $N$, when the MF contributions are no longer suppressed, the beyond-MF term becomes negligible. 

Although beyond-MF corrections do not scale up with $N$, we find cases where they can be relevant even for large systems because of a strong suppression of the total shift by the multilevel structure. An example is the case with a pulse area of $\theta=\pi/4$ presented in Fig.~\ref{Fig:shift_BMF_scaling}(b) and (d) for the same 2D lattice configuration of (a) and (c). Due to the strong suppression of the total shift when addressing the $-1/2$ transition [see Eq.~(\ref{Eq:zero_order_shift})], the beyond-MF contributions become comparable in magnitude to the actual shift. Figure~\ref{Fig:shift_BMF_scaling}(b) shows that this holds true for lattices of size up to $\sim 10^3$ atoms. Note, however, that in this case the beyond-MF term is suppressed in the large B-field limit.

%###################################################################%
%######################      CONCLUSION     ########################%
%###################################################################%

\textit{Conclusion.---}We have shown that dipolar frequency shifts are strongly modified in systems featuring a multilevel structure. The predicted two orders of magnitude suppression obtained by properly addressing specific transitions can lead to the improved accuracy necessary for the exploration of fundamental physics~\cite{Chou2010,Grotti2018,Sanner2019,Derevianko2014,Kennedy2020}, providing new insights on the behavior of strongly and long-range interacting many-body systems.

%###################################################################%
%###################      ACKNOWLEDGEMENTS     #####################%
%###################################################################%
\begin{acknowledgments}
\textit{Acknowledgments.---}We thank C. Qu, L. Sonderhouse, and N. Schine for helpful discussions and feedback. A.C. and R.B. are supported by FAPESP through Grants No. 2017/09390-7, 2018/18353-0, 2019/13143-0, and 2018/15554-5. R.B. benefited from Grants from the National Council for Scientific and Technological Development (CNPq, Grant Nos. 302981/2017-9 and 409946/2018-4). C.S. thanks the Humboldt Foundation for support. This work is supported by the AFOSR Grant No. FA9550-18-1-0319 and its MURI Initiative, by the DARPA and ARO Grant No. W911NF-16-1-0576, the ARO single investigator Grant No. W911NF-19-1-0210, the NSF PHY1820885, NSF JILA-PFC PHY-1734006 Grants, NSF QLCI-2016244 grant, and by NIST.
\end{acknowledgments}

\bibliography{bib}

\pagebreak

\clearpage

%####################################################################%
%##############      SUPPLEMENTAL MATERIAL     ######################%
%####################################################################%
%%%%%%%%%% Prefix a "S" to all equations, figures, tables and reset the counter %%%%%%%%%%
\setcounter{equation}{0}
\setcounter{figure}{0}
\setcounter{table}{0}
\setcounter{page}{1}
\makeatletter
\renewcommand{\theequation}{S\arabic{equation}}
\renewcommand{\thefigure}{S\arabic{figure}}
\renewcommand{\bibnumfmt}[1]{[S#1]}
\renewcommand{\citenumfont}[1]{S#1}
\onecolumngrid
\vspace{0.5cm}
\begin{center}
\textbf{\large Supplemental material: \papertitle}\\[.2cm]
\end{center}

% ----- Definitions for the Supplemental Material 
\newcommand{\ma}{\alpha}
\newcommand{\mb}{\gamma}
\newcommand{\mc}{\eta}
\newcommand{\md}{\zeta}
\newcommand{\ga}{g_\ma}
\newcommand{\gb}{g_\mb}
\newcommand{\gc}{g_\mc}
\newcommand{\gd}{g_\md}

\newcommand{\na}{\beta}
\newcommand{\nb}{\gamma}
\newcommand{\nc}{\eta}
\newcommand{\nd}{\zeta}
\newcommand{\ea}{e_\na}
\newcommand{\eb}{e_\nb}
\newcommand{\ec}{e_\nc}
\newcommand{\ed}{e_\nd}

\newcommand{\m}{m}
\newcommand{\mm}{m'}
\newcommand{\n}{n}
\newcommand{\nn}{n'}

\newcommand{\G}{\mathcal{G}}

% Useful color code for labels in equations 

\newcommand{\ca}[1]{{\textcolor{blue}#1}}
\newcommand{\cb}[1]{{\textcolor{red}#1}}
\newcommand{\cc}[1]{{\textcolor{violet}#1}}
\newcommand{\cd}[1]{{\textcolor{teal}#1}}

%##################################################################%	
%------------ 	First-order frequency shift: definitions ----------%
%##################################################################%	
\section{First-order frequency shift: definitions}
The coefficients for the short-time expansion of the dipolar shift  [Eqs.~(\ref{Eq:zero_order_shift}) and (\ref{Eq:first_order_shift}) in the main text] are given by $\delta^{\g\e}(t)\approx \delta^{\g\e}_0 + \delta^{\g\e}_1 t$ with
\Eq{Eq:zero_order_shift_SM}{\delta_0^{\g\e}&=\frac{1}{2 \pi}\frac{\ehS{y}_1}{\ehS{x}_0}= -\frac{\cos\theta}{2\pi N}\sum\limits_{i,j\neq i}U^{ji}_{\g\e},}
and
\Eq{Eq:first_order_shift_SM}{\delta^{\g\e}_1& =\frac{1}{4\pi}\left(\frac{\ehS{y}_2}{\ehS{x}_0}-2\frac{\ehS{x}_1\ehS{y}_1}{\ehS{x}_0^2}\right)=-\frac{1}{2\pi N}\sum_{i,j\neq i}\Bigg\{U^{ji}_{\g\e}\widetilde{\Gamma}_{\g\e}\left(\theta\right)+\sum_{p}\Bigg(\sum_{k\neq i,j}W^{kji}_p\left(\theta\right)+\sum_{p'}Q^{ji}_{p,p'}\left(\theta\right)\Bigg)\Bigg\}.
}
Recall that we expand expectation values of operators as $\langle\hat{\mathcal{O}}\rangle \approx \langle\hat{\mathcal{O}}\rangle_0 + \langle\hat{\mathcal{O}}\rangle_1 t + \langle\hat{\mathcal{O}}\rangle_2 t^2/2$.
In the following, we provide the detailed definitions of the terms contained in the above expressions, starting with
\Eq{Eq:U_and_E_def}{& U^{ji}_{g_me_n,g_{m'}e_{n'}}(\br_{ik})\equiv \Gamma^{ji}_{g_me_n,g_{m'}e_{n'}}\sin(\kr{ik})+ \Delta^{ji}_{g_me_n,g_{m'}e_{n'}}\cos(\kr{ik}),\\
& E^{ji}_{g_me_n,g_{m'}e_{n'}}(\br_{ik})\equiv \Gamma^{ji}_{g_me_n,g_{m'}e_{n'}}\cos(\kr{ik})- \Delta^{ji}_{g_me_n,g_{m'}e_{n'}}\sin(\kr{ik}).}

\noindent For the sake of simplicity, we have further defined (as used in the main text)
\Eq{Eq:U_and_E_def_transition}{U^{ji}_{\g\e}\equiv U^{ji}_{\g\e,\g\e}(\br_{ij}) \text{ and } E^{ji}_{\g\e}\equiv E^{ji}_{\g\e,\g\e}(\br_{ij}).}

\noindent The first-order shift contains the term
\Eq{Eq:wide_tilde_Gamma}{\widetilde{\Gamma}_{\g\e}\left(\theta\right)\equiv\sin^2(\theta/2)(1+|C_{\g}^{\e}|^2)\frac{\Gamma}{2}+(\cos^2\theta/N)\sum_{i,j\neq i}E^{ji}_{\g\e}.}

\noindent The two-body terms that involve two-photon transitions between levels stem exclusively from beyond-MF contributions and are given by
\Eq{Eq:Q_def}{
Q^{ji}_{p,p'}\left(\theta\right) \equiv \frac{1}{2}\left(\sin^2(\theta/2)A^{- ji;ij}_{g_{\me+p}\e,g_{\me+p'}\e}-\cos^2(\theta/2)A^{- ji;ij}_{\g e_{\mg-p'},g_{\me+p}\e}\right),}
while the three-body terms read
\Eq{Eq:W_def}{W^{kji}_{p}\left(\theta\right)\equiv
        &\frac{\sin^2\theta}{8}
                \Big(V^{- ji;ik}_{\g e_{\mg-p},\g\e}+ V^{- ji;ik}_{g_{\me+p}\e,\g\e} + 2\delta_{p,(\mg-\me)} V^{+ ji;ik}_{\g\e,\g\e}\Big)
            \\ 
        -&\frac{\cos\theta}{2}
                \Big(\cos^2(\theta/2) D^{ij;jk}_{\g e_{\mg-p},\g\e}- \sin^2(\theta/2) D^{ij;jk}_{g_{\me+p}\e,\g\e}\Big),
}
with
\Eq{Eq:V_def}{V^{\pm ji;ik}_{g_me_n,g_{m'}e_{n'}}\equiv A^{\mp ji;ik}_{g_me_n,g_{m'}e_{n'}}a^{\mp}(\br_{ij},\br_{ik})+B^{\pm ji;ik}_{g_me_n,g_{m'}e_{n'}} b^{\pm}(\br_{ij},\br_{ik}),}
\Eq{Eq:A_and_B_def}{& A^{\pm ji;ik}_{g_me_n,g_{m'}e_{n'}}\equiv \Delta^{ji}_{g_me_n,g_{m'}e_{n'}}\Gamma^{ik}_{g_{m'}e_{n'},g_me_n}\mp\Gamma^{ji}_{g_me_n,g_{m'}e_{n'}}\Delta^{ik}_{g_{m'}e_{n'},g_me_n},\\
& B^{\pm ji;ik}_{g_me_n,g_{m'}e_{n'}}\equiv \Delta^{ji}_{g_me_n,g_{m'}e_{n'}}\Delta^{ik}_{g_{m'}e_{n'},g_me_n}\mp\Gamma^{ji}_{g_me_n,g_{m'}e_{n'}}\Gamma^{ik}_{g_{m'}e_{n'},g_me_n},
}
\Eq{Eq:trig_def}{
& a^{\pm}(\br_{ij},\br_{ik})\equiv \cos(\kr{ij})\cos(\kr{ik})\pm\sin(\kr{ij})\sin(\kr{ik}),\\
& b^{\pm}(\br_{ij},\br_{ik})\equiv \sin(\kr{ij})\cos(\kr{ik})\pm\cos(\kr{ij})\sin(\kr{ik}),
}
and 
\Eq{Eq:D_def}{& D^{ij;jk}_{g_me_n,g_{m'}e_{n'}} \equiv \Delta^{ij}_{g_me_n,g_{m'}e_{n'}}E^{jk}_{g_{m'}e_{n'},g_me_n}\left(\br_{ik}\right)+\Gamma^{ij}_{g_me_n,g_{m'}e_{n'}}U^{jk}_{g_{m'}e_{n'},g_me_n}\left(\br_{ik}\right).}

\noindent As mentioned in the main text, the dipolar shift computed in MF approximation deviates from the exact result at first order in time. Compared to Eq.~(\ref{Eq:first_order_shift_SM}), the MF expression does not have the $Q^{ji}_{p,p'}$ terms and it contains an additional term which is equal to the three-body term of Eq.~(\ref{Eq:W_def}) after setting $k=j$ and $k=i$ in the first and second lines, respectively. Specifically, the MF shift is given by $\delta^{\g\e}_{0,\text{MF}}=\delta^{\g\e}_{0}$ and
\Eq{Eq:first_order_shift_MF_SM}{
\delta^{\g\e}_{1,\text{MF}}& =-\frac{1}{2\pi N}\sum_{i,j\neq i}\Bigg\{U^{ji}_{\g\e}\widetilde{\Gamma}_{\g\e}\left(\theta\right)+\sum_{p}\Bigg( W^{ji}_{\text{MF},p}\left(\theta\right) + \sum_{k\neq i,j}W^{kji}_p\left(\theta\right)\Bigg)\Bigg\},
}
with
\Eq{Eq:W_MF_def}{W^{ji}_{\text{MF},p}\left(\theta\right)\equiv
        \frac{\sin^2\theta}{4}\delta_{p,(\mg-\me)} V^{+ ji;ij}_{\g\e,\g\e} 
        -\frac{\cos\theta}{2}\Big(\cos^2(\theta/2) D^{ij;ji}_{\g e_{\mg-p},\g\e}- \sin^2(\theta/2) D^{ij;ji}_{g_{\me+p}\e,\g\e}\Big).
}

%###################################################################%	
%-------------------- Equations of motion --------------------------%
%###################################################################%

\section{Equations of motion}

Here we provide the equations of motion for the expectation values derived from the multilevel coupled dipole master equation introduced in the main text. We also include in the equations the Zeeman shifts induced by a magnetic field parallel to the quantization axis. The Zeeman Hamiltonian is given by $\hat{H}_{B}=-\sum_{n,i} \Delta_{e_n} \hat{\sigma}_i^{e_ne_n} - \sum_{m,i} \Delta_{g_m} \hat{\sigma}_i^{g_mg_m}$, where $\Delta_{e_n}=n\delta_e$, $\Delta_{g_m}=m\delta_g$, and $\delta_g\neq\delta_e$ due to a differential $g$-factor between ground and excited manifolds.

We use Einstein notation for sums over level indices (i.e., repeated indices $m,m',n,n'$ are summed if they do not appear on the left hand side of the equation). Additionally, note that for the sake of simplicity in notation we have chosen to simplify $\G^{ij}_{g_m e_{n},g_{m'} e_{n'}} \rightarrow  \G^{ij}_{mn,m'n'}$.

The single-point equations read
%-------------------------------------------------------------------%
%%%%%%%%%%%%%%%% SINGLE-OPERATOR EXPECTATION VALUES %%%%%%%%%%%%%%%%%
%-------------------------------------------------------------------%
 
% ---------------- EQUATION seg: 
% OBS: use the redefined labels below to write equation indices accordingly
% --- Variables indices
\renewcommand{\ma}{\cb{\alpha}}
\renewcommand{\mb}{\gamma}
\renewcommand{\mc}{\eta}
\renewcommand{\na}{\ca{\beta}}
\renewcommand{\nb}{\gamma}
\renewcommand{\nc}{\eta}
% --- Summation indices
\renewcommand{\m}{m}
\renewcommand{\mm}{{m'}}
\renewcommand{\n}{n}
\renewcommand{\nn}{{n'}}
% ---- Atom index
\renewcommand{\i}{i}
\renewcommand{\j}{j}
\renewcommand{\k}{k}

\Eq{Eq:seg_mag}{\frac{d}{dt}\ehs{\ea \ga}{\i}= & -\Bigg(\frac{\Gamma}{2}+i(\Delta_{\ea}-\Delta_{\ga})\Bigg)\ehs{\ea \ga}{\i}+\sum_{\j\neq \i}\left(\G^{\j\i}_{\m\n,\ma \nn}\ehss{e_\n g_\m}{\j}{\ea e_\nn}{\i}-\G^{\j\i}_{\m\n,\mm\na}\ehss{e_\n g_\m}{\j}{g_\mm \ga}{\i}\right),}

% ---------------- EQUATION see: 
% OBS: use the redefined labels below to write equation indices accordingly
% --- Variables indices
\renewcommand{\ma}{\alpha}
\renewcommand{\mb}{\gamma}
\renewcommand{\mc}{\eta}
\renewcommand{\na}{\ca{\beta}}
\renewcommand{\nb}{\cb{\gamma}}
\renewcommand{\nc}{\eta}
% --- Summation indices
\renewcommand{\m}{m}
\renewcommand{\mm}{{m'}}
\renewcommand{\n}{n}
\renewcommand{\nn}{{n'}}
% ---- Atom index
\renewcommand{\i}{i}
\renewcommand{\j}{j}
\renewcommand{\k}{k}

\Eq{Eq:see_mag}{\frac{d}{dt}\ehs{\ea \eb}{\i} & = -i(\Delta_{\ea}-\Delta_{\eb})\ehs{\ea \eb}{\i}
-\left(\G^{*\i\i}_{\m\nb,\m\n}\ehs{\ea e_\n}{\i}+\G^{\i\i}_{\m\n,\m\na}\ehs{e_\n \eb}{\i}\right)
\\
-&\sum_{\j\neq \i}\left(\G^{*\i\j}_{\mm\nb,\m\n}\ehss{\ea g_\mm}{\i}{g_\m e_\n}{\j}+\G^{\j\i}_{\m\n,\mm\na}\ehss{e_\n g_\m}{\j}{g_\mm\eb}{\i}\right),}

% ---------------- EQUATION sgg: 
% OBS: use the redefined labels below to write equation indices accordingly
% --- Variables indices
\renewcommand{\ma}{{\textcolor{blue}\alpha}}
\renewcommand{\mb}{{\textcolor{red}\gamma}}
\renewcommand{\mc}{\eta}
\renewcommand{\na}{\beta}
\renewcommand{\nb}{\gamma}
\renewcommand{\nc}{\eta}
% --- Summation indices
\renewcommand{\m}{m}
\renewcommand{\mm}{{m'}}
\renewcommand{\n}{n}
\renewcommand{\nn}{{n'}}
% ---- Atom index
\renewcommand{\i}{i}
\renewcommand{\j}{j}
\renewcommand{\k}{k}

\Eq{Eq:sgg_mag}{\frac{d}{dt}\ehs{\ga \gb}{\i} & = -i(\Delta_{\ga}-\Delta_{\gb})\ehs{\ga \gb}{\i}
+\left(\G^{*\i\i}_{\ma\nn,\mb\n}+\G^{\i\i}_{\ma\nn,\mb\n}\right)\ehs{e_\n e_\nn}{\i}
\\
+&\sum_{\j\neq \i}\left(\G^{\j\i}_{\m\nn,\mb\n}\ehss{e_\nn g_\m}{\j}{g_\ma e_\n}{\i}+\G^{*\i\j}_{\ma\nn,\m\n}\ehss{e_\nn g_\mb}{\i}{g_\m e_\n}{\j}\right).}

The two-point equations read 
%-------------------------------------------------------------------%
%%%%%%%%%%%%%%%% TWO-OPERATOR EXPECTATION VALUES %%%%%%%%%%%%%%%%%
%-------------------------------------------------------------------%

% ---------------- EQUATION seg_see: 
% OBS: use the redefined labels below to write equation indices accordingly

\renewcommand{\ma}{\cb{\alpha}}
\renewcommand{\mb}{{\gamma}}
\renewcommand{\mc}{{\eta}}
\renewcommand{\md}{\zeta}
\renewcommand{\na}{\ca{\beta}}
\renewcommand{\nb}{\cc{\gamma}}
\renewcommand{\nc}{\cd{\eta}}
\renewcommand{\nd}{\zeta}
% --- Summation indices
\renewcommand{\m}{m}
\renewcommand{\mm}{{m'}}
\renewcommand{\n}{n}
\renewcommand{\nn}{{n'}}
% ---- Atom index
\renewcommand{\i}{i}
\renewcommand{\j}{j}
\renewcommand{\k}{k}

% UNCOMMENT BELOW IF YOU WANT THE INDICES TO BE STRAIGHTFORWARDLY THE ONES WE
% USE FOR THE SHIFT CALCULATION

% --- Variables indices
% \renewcommand{\ma}{m}
% \renewcommand{\mb}{\gamma}
% \renewcommand{\mc}{\eta}
% \renewcommand{\na}{n}
% \renewcommand{\nb}{\beta}
% \renewcommand{\nc}{{n'}}
% % --- Summation indices
% \renewcommand{\m}{l}
% \renewcommand{\mm}{{l'}}
% \renewcommand{\n}{q}
% \renewcommand{\nn}{{q'}}
% % ---- Atom index
% \renewcommand{\i}{j}
% \renewcommand{\j}{i}
% \renewcommand{\k}{k}

\Eq{Eq:seg_see_mag}{\frac{d}{dt}\ehss{\ea \ga}{\i}{\eb \ec}{\j} & = -i(\Delta_{\ea} -\Delta_{\ga} + \Delta_{\eb} -\Delta_{\ec})\ehss{\ea \ga}{\i}{\eb \ec}{\j}\\
& -\Big( \begin{aligned}[t] & \G^{*\j\i}_{\m \nc,\ma \n}\ehss{\ea e_\n}{\i}{\eb g_\m}{\j} + \G^{\i\i}_{\m\n,\m\na}\ehss{e_\n \ga}{\i}{\eb \ec}{\j}\\
+& \G^{*\j\j}_{\m\nc,\m\n}\ehss{\ea \ga}{\i}{\eb e_\n}{\j} + \G^{\j\j}_{\m\n,\m\nb}\ehss{\ea \ga}{\i}{e_\n\ec}{\j}\Big)
\end{aligned}\\
&-\sum_{\k\neq \i,\j}\begin{aligned}[t]\Big( & \G^{*\j\k}_{\mm\nc,\m\n}\ehsss{\ea \ga}{\i}{\eb g_{\mm}}{\j}{g_\m e_\n}{\k} + \G^{\k\i}_{\mm\n,\m\na}\ehsss{g_\m \ga}{\i}{\eb \ec}{\j}{e_\n g_\mm}{\k}\\
+ & \G^{\k\j}_{\mm\n,\m\nb}\ehsss{\ea \ga}{\i}{g_\m \ec}{\j}{e_\n g_\mm}{\k} - \G^{\k\i}_{\m \nn,\ma \n}\ehsss{\ea e_\n}{\i}{\eb \ec}{\j}{e_\nn g_\m}{\k} \Big),\end{aligned}}

% ---------------- EQUATION seg_sgg: 
% OBS: use the redefined labels below to write equation indices accordingly
% --- Variables indices

\renewcommand{\ma}{\cb{\alpha}}
\renewcommand{\mb}{\cc{\gamma}}
\renewcommand{\mc}{\cd{\eta}}
\renewcommand{\md}{\zeta}
\renewcommand{\na}{\ca{\beta}}
\renewcommand{\nb}{\gamma}
\renewcommand{\nc}{\eta}
\renewcommand{\nd}{\zeta}
% --- Summation indices
\renewcommand{\m}{m}
\renewcommand{\mm}{{m'}}
\renewcommand{\n}{n}
\renewcommand{\nn}{{n'}}
% ---- Atom index
\renewcommand{\i}{i}
\renewcommand{\j}{j}
\renewcommand{\k}{k}

% UNCOMMENT BELOW IF YOU WANT THE INDICES TO BE STRAIGHTFORWARDLY THE ONES WE
% USE FOR THE SHIFT CALCULATION

% \renewcommand{\ma}{m}
% \renewcommand{\mb}{{m'}}
% \renewcommand{\mc}{\alpha}
% \renewcommand{\na}{n}
% \renewcommand{\nb}{\gamma}
% \renewcommand{\nc}{\eta}
% % --- Summation indices
% \renewcommand{\m}{l}
% \renewcommand{\mm}{{l'}}
% \renewcommand{\n}{q}
% \renewcommand{\nn}{{q'}}
% % ---- Atom index
% \renewcommand{\i}{j}
% \renewcommand{\j}{i}
% \renewcommand{\k}{k}

\Eq{Eq:seg_sgg_mag}{\frac{d}{dt}\ehss{\ea \ga}{\i}{\gb \gc}{\j} & =-i(\Delta_{\ea} -\Delta_{\ga} + \Delta_{\gb} -\Delta_{\gc})\ehss{\ea \ga}{\i}{\gb \gc}{\j}\\
& +\Big(\G^{\j\j}_{\mb \nn,\mc \n}+\G^{*\j\j}_{\mb \nn,\mc \n}\Big)\ehss{\ea \ga}{\i}{e_\nn e_\n}{\j}+ \Big(\G^{\j\i}_{\mb \nn,\ma \n}+\G^{*\j\i}_{\mb \nn,\ma \n}\Big)\ehss{\ea e_\n}{\i}{e_\nn \gc}{\j}\\
&- \Big(\G^{\i\i}_{\m \n,\m\na}\ehss{e_\n \ga}{\i}{\gb \gc}{\j} + \G^{\j\i}_{\mb \n,\m\na}\ehss{g_\m \ga}{\i}{e_\n \gc}{\j}\Big)\\
&+\sum_{\k\neq \i,\j}\begin{aligned}[t]\Big( & \G^{\k\j}_{\m\n,\mc \nn}\ehsss{\ea \ga}{\i}{\gb e_\nn}{\j}{e_\n g_\m}{\k} + \G^{*\j\k}_{\mb \nn,\m\n}\ehsss{\ea \ga}{\i}{e_{\nn}\gc}{\j}{g_\m e_\n}{\k}\\
+ & \G^{\k\i}_{\m\n,\ma \nn}\ehsss{\ea e_\nn}{\i}{\gb \gc}{\j}{e_\n g_\m}{\k} - \G^{\k\i}_{\m\n,\mm\na}\ehsss{g_\mm \ga}{\i}{\gb \gc}{\j}{e_\n g_\m}{\k} \Big),\end{aligned}}

% ---------------- EQUATION see_see: 
% OBS: use the redefined labels below to write equation indices accordingly
% --- Variables indices
\renewcommand{\ma}{\alpha}
\renewcommand{\mb}{\gamma}
\renewcommand{\mc}{\eta}
\renewcommand{\md}{\zeta}
\renewcommand{\na}{{\textcolor{blue}\beta}}
\renewcommand{\nb}{{\textcolor{red}\gamma}}
\renewcommand{\nc}{{\textcolor{violet}\eta}}
\renewcommand{\nd}{{\textcolor{teal}\zeta}}
% --- Summation indices
\renewcommand{\m}{m}
\renewcommand{\mm}{{m'}}
\renewcommand{\n}{n}
\renewcommand{\nn}{{n'}}
% ---- Atom index
\renewcommand{\i}{i}
\renewcommand{\j}{j}
\renewcommand{\k}{k}

\Eq{Eq:see_see_mag}{\frac{d}{dt}\ehss{\ea\eb}{\i}{\ec\ed}{\j} & = -i(\Delta_{\ea} -\Delta_{\eb} + \Delta_{\ec} -\Delta_{\ed})\ehss{\ea\eb}{\i}{\ec\ed}{\j}\\
& -\Big( \begin{aligned}[t] & \G^{*\i\i}_{\m\nb,\m\n}\ehss{\ea e_\n}{\i}{\ec\ed}{\j} + \G^{\i\i}_{\m\n,\m\na}\ehss{e_\n\eb}{\i}{\ec\ed}{\j}\\
+& \G^{*\j\j}_{\m\nd,\m\n}\ehss{\ea\eb}{\i}{\ec e_\n}{\j} + \G^{\j\j}_{\m\n,\m\nc}\ehss{\ea\eb}{\i}{e_\n\ed}{\j}\Big)
\end{aligned}\\
&-\sum_{\k\neq \i,\j}\begin{aligned}[t]\Big( & 
\G^{*\j\k}_{\mm\nd,\m\n}\ehsss{\ea\eb}{\i}{\ec g_\mm}{\j}{ g_\m e_\n}{\k} 
+ \G^{\k\j}_{\mm\n,\m\nc}\ehsss{\ea\eb}{\i}{g_\m\ed}{\j}{e_\n g_\mm}{\k}\\
+ & \G^{*\i\k}_{\mm \nb,\m\n}\ehsss{\ea g_\mm}{\i}{\ec\ed}{\j}{g_\m e_\n}{\k}  
+ \G^{\k\i}_{\mm\n,\m\ea}\ehsss{g_\m \eb}{\i}{\ec\ed}{\j}{e_\n g_\mm}{\k} \Big),\end{aligned}}

% ---------------- EQUATION sgg_sgg: 
% OBS: use the redefined labels below to write equation indices accordingly
% --- Variables indices
% --- Variables indices
\renewcommand{\ma}{{\textcolor{blue}\alpha}}
\renewcommand{\mb}{{\textcolor{red}\gamma}}
\renewcommand{\mc}{{\textcolor{violet}\eta}}
\renewcommand{\md}{{\textcolor{teal}\zeta}}
\renewcommand{\na}{{\beta}}
\renewcommand{\nb}{{\gamma}}
\renewcommand{\nc}{{\eta}}
\renewcommand{\nd}{{\zeta}}
% --- Summation indices
\renewcommand{\m}{m}
\renewcommand{\mm}{{m'}}
\renewcommand{\n}{n}
\renewcommand{\nn}{{n'}}
% ---- Atom index
\renewcommand{\i}{i}
\renewcommand{\j}{j}
\renewcommand{\k}{k}

\Eq{Eq:sgg_sgg_mag}{\frac{d}{dt}\ehss{\ga\gb}{\i}{\gc\gd}{\j} & = -i(\Delta_{\ga} -\Delta_{\gb} + \Delta_{\gc} -\Delta_{\gd})\ehss{\ga\gb}{\i}{\gc\gd}{\j}\\
& +\Big[ \begin{aligned}[t] 
&  \left(\G^{\i\j}_{\ma\nn,\md\n}+\G^{*\i\j}_{\ma\nn,\md\n}\right)\ehss{e_\nn \gb}{\i}{\gc e_\n}{\j} \\
+& \left(\G^{\i\i}_{\ma\nn,\mb\n}+\G^{*\i\i}_{\ma\nn,\mb\n}\right)\ehss{e_\nn e_\n}{\i}{\gc\gd}{\j}\\
+& \left(\G^{\j\j}_{\mc\nn,\md\n}+\G^{*\j\j}_{\mc\nn,\md\n}\right)\ehss{\ga\gb}{\i}{e_\nn e_\n}{\j} \\
+& \left(\G^{\j\i}_{\mc\nn,\mb\n}+\G^{*\j\i}_{\mc\nn,\mb\n}\right)\ehss{\ga e_\n}{\i}{e_\nn\gd}{\j} \Big]
\end{aligned}\\
&+\sum_{\k\neq \i,\j}\begin{aligned}[t]\Big( & \G^{\k\j}_{\m\nn,\md\n}\ehsss{\ga\gb}{\i}{\gc e_\n}{\j}{e_\nn g_\m}{\k} + \G^{*\j\k}_{\mc\nn,\m\n}\ehsss{\ga\gb}{\i}{e_\nn\gd}{\j}{g_\m e_\n}{\k}\\
+ & \G^{\k\i}_{\m\nn,\mb\n}\ehsss{\ga e_\n}{\i}{\gc\gd}{\j}{e_\nn g_\m}{\k} + \G^{*\i\k}_{\ma\nn,\m\n}\ehsss{e_\nn\gb}{\i}{\gc\gd}{\j}{g_\m e_\n}{\k} \Big),\end{aligned}}

% ---------------- EQUATION see_sgg: 
% OBS: use the redefined labels below to write equation indices accordingly
% --- Variables indices
% --- Variables indices
\renewcommand{\ma}{\cc{\alpha}}
\renewcommand{\mb}{\cd{\eta}}
\renewcommand{\mc}{\eta}
\renewcommand{\md}{\zeta}
\renewcommand{\na}{\ca{\beta}}
\renewcommand{\nb}{\cb{\gamma}}
\renewcommand{\nc}{{\eta}}
\renewcommand{\nd}{{\zeta}}
% --- Summation indices
\renewcommand{\m}{m}
\renewcommand{\mm}{{m'}}
\renewcommand{\n}{n}
\renewcommand{\nn}{{n'}}
% ---- Atom index
\renewcommand{\i}{i}
\renewcommand{\j}{j}
\renewcommand{\k}{k}

\Eq{Eq:see_sgg_mag}{\frac{d}{dt}\ehss{\ea\eb}{\i}{\ga\gb}{\j} & = -i(\Delta_{\ea} -\Delta_{\eb} + \Delta_{\ga} -\Delta_{\gb})\ehss{\ea\eb}{\i}{\ga\gb}{\j}\\
& +\Big[ \begin{aligned}[t] 
 &\left(\G^{\j\j}_{\ma\nn,\mb\n} + \G^{*\j\j}_{\ma\nn,\mb\n}\right)\ehss{\ea\eb}{\i}{e_\nn e_\n}{\j}\\
-\Big(&\G^{*\i\i}_{\m\nb,\m\n}\ehss{\ea e_\n}{\i}{\ga\gb}{\j} + \G^{\i\i}_{\m\n,\m\na}\ehss{e_\n \eb}{\i}{\ga\gb}{\j}\\
        +&\G^{*\i\j}_{\m\nb,\mb\n}\ehss{\ea g_\m}{\i}{\ga e_\n}{\j} + \G^{\j\i}_{\ma\n,\m\na}\ehss{g_\m \eb}{\i}{e_\n\gb}{\j} \Big)\Big]
\end{aligned}\\
&+\sum_{\k\neq \i,\j}\begin{aligned}[t]\Big( & \G^{\k\j}_{\m\nn,\mb\n}\ehsss{\ea\eb}{\i}{\ga e_\n}{\j}{e_\nn g_\m}{\k} + \G^{*\j\k}_{\ma\nn,\m\n}\ehsss{\ea\eb}{\i}{e_\nn \gb}{\j}{g_\m e_\n}{\k}\\
- & \G^{*\i\k}_{\mm\nb,\m\n}\ehsss{\ea g_\mm}{\i}{\ga\gb}{\j}{g_\m e_\n}{\k} - \G^{\k\i}_{\mm\n,\m\na}\ehsss{g_\m \eb}{\i}{\ga\gb}{\j}{e_\n g_\mm}{\k} \Big),\end{aligned}}

% ---------------- EQUATION seg_sge: 
% OBS: use the redefined labels below to write equation indices accordingly
% --- Variables indices
% --- Variables indices
\renewcommand{\ma}{\cb{\alpha}}
\renewcommand{\mb}{\cc{\eta}}
\renewcommand{\mc}{\eta}
\renewcommand{\md}{\zeta}
\renewcommand{\na}{\ca{\beta}}
\renewcommand{\nb}{\cd{\gamma}}
\renewcommand{\nc}{{\eta}}
\renewcommand{\nd}{{\zeta}}
% --- Summation indices
\renewcommand{\m}{m}
\renewcommand{\mm}{{m'}}
\renewcommand{\n}{n}
\renewcommand{\nn}{{n'}}
% ---- Atom index
\renewcommand{\i}{i}
\renewcommand{\j}{j}
\renewcommand{\k}{k}

\Eq{Eq:seg_sge_mag}{\frac{d}{dt}\ehss{\ea\ga}{\i}{\gb\eb}{\j} & = -i(\Delta_{\ea} -\Delta_{\ga} + \Delta_{\gb} -\Delta_{\eb})\ehss{\ea\ga}{\i}{\gb\eb}{\j}\\
& +\Big[ \begin{aligned}[t] 
 &\left(\G^{\j\i}_{\mb\nn,\ma\n} + \G^{*\j\i}_{\mb\nn,\ma\n}\right)\ehss{\ea e_\n}{\i}{e_\nn\eb}{\j}\\
-\Big(&\G^{*\j\i}_{\m\nb,\ma\n}\ehss{\ea e_\n}{\i}{\gb g_\m}{\j} + \G^{\i\i}_{\m\n,\m\na}\ehss{e_\n \ga}{\i}{\gb\eb}{\j}\\
        +&\G^{*\j\j}_{\m\nb,\m\n}\ehss{\ea \ga}{\i}{\gb e_\n}{\j} + \G^{\j\i}_{\mb\n,\m\na}\ehss{g_\m \ga}{\i}{e_\n\eb}{\j} \Big)\Big]
\end{aligned}\\
&+\sum_{\k\neq \i,\j}\begin{aligned}[t]\Big( & \G^{*\j\k}_{\mb\nn,\m\n}\ehsss{\ea\ga}{\i}{e_\nn\eb}{\j}{g_\m e_\n}{\k} + \G^{\k\i}_{\m\nn,\ma\n}\ehsss{\ea e_\n}{\i}{\gb \eb}{\j}{e_\nn g_\m}{\k}\\
- & \G^{*\j\k}_{\mm\nb,\m\n}\ehsss{\ea\ga}{\i}{\gb g_\mm}{\j}{g_\m e_\n}{\k} - \G^{\k\i}_{\mm\n,\m\na}\ehsss{g_\m \ga}{\i}{\gb\eb}{\j}{e_\n g_\mm}{\k} \Big),\end{aligned}}

% ---------------- EQUATION seg_seg: 
% OBS: use the redefined labels below to write equation indices accordingly
% --- Variables indices
% --- Variables indices
\renewcommand{\ma}{\cb{\alpha}}
\renewcommand{\mb}{\cd{\eta}}
\renewcommand{\mc}{\eta}
\renewcommand{\md}{\zeta}
\renewcommand{\na}{\ca{\beta}}
\renewcommand{\nb}{\cc{\gamma}}
\renewcommand{\nc}{{\eta}}
\renewcommand{\nd}{{\zeta}}
% --- Summation indices
\renewcommand{\m}{m}
\renewcommand{\mm}{{m'}}
\renewcommand{\n}{n}
\renewcommand{\nn}{{n'}}
% ---- Atom index
\renewcommand{\i}{i}
\renewcommand{\j}{j}
\renewcommand{\k}{k}

\Eq{Eq:seg_seg_mag}{\frac{d}{dt}\ehss{\ea\ga}{\i}{\eb\gb}{\j} & =  -i(\Delta_{\ea} -\Delta_{\ga} + \Delta_{\eb} -\Delta_{\gb})\ehss{\ea\ga}{\i}{\eb\gb}{\j}\\
& \begin{aligned}[t] 
-\Big(&\G^{\i\i}_{\m\n,\m\na}\ehss{e_\n \ga}{\i}{\eb \gb}{\j} + \G^{\j\j}_{\m\n,\m\nb}\ehss{\ea \ga}{\i}{e_\n\gb}{\j} \Big)
\end{aligned}\\
&+\sum_{\k\neq \i,\j}\begin{aligned}[t]\Big( & \G^{\k\j}_{\m\nn,\mb\n}\ehsss{\ea\ga}{\i}{\eb e_\n}{\j}{e_\nn g_\m}{\k} + \G^{\k\i}_{\m\nn,\ma\n}\ehsss{\ea e_\n}{\i}{\eb \gb}{\j}{e_\nn g_\m}{\k}\\
- & \G^{\k\j}_{\mm\n,\m\nb}\ehsss{\ea\ga}{\i}{g_\m\gb}{\j}{e_\n g_\mm }{\k} - \G^{\k\i}_{\mm\n,\m\na}\ehsss{g_\m \ga}{\i}{\eb\gb}{\j}{e_\n g_\mm}{\k} \Big).\end{aligned}}

We have defined $\G^{ij}_{mn,m'n'}\equiv\Gamma^{ij}_{mn,m'n'}+i\Delta^{ij}_{mn,m'n'}$ and $\G^{*ij}_{mn,m'n'}\equiv\Gamma^{ij}_{mn,m'n'}-i\Delta^{ij}_{mn,m'n'}$.
%###################################################################%	
%----------- Magnetic fields in the short-time expansion -----------%
%###################################################################%

\section{Short-time expansion with magnetic fields}

Here we provide details of the short-time perturbative expansion in the presence of a magnetic field. We show (1) that the first-order term of the dipolar shift is independent from Zeeman shifts of any size, and (2) we derive simplified expressions for the shift in the large Zeeman shift  limit.

% ---------------- EQUATION seg: 
% OBS: use the redefined labels below to write equation indices accordingly
% --- Variables indices
\renewcommand{\ma}{\alpha}
\renewcommand{\mb}{\gamma}
\renewcommand{\mc}{\eta}
\renewcommand{\na}{\beta}
\renewcommand{\nb}{\gamma}
\renewcommand{\nc}{\eta}
% --- Summation indices
\renewcommand{\m}{m}
\renewcommand{\mm}{{m'}}
\renewcommand{\n}{n}
\renewcommand{\nn}{{n'}}
% ---- Atom index
\renewcommand{\i}{i}
\renewcommand{\j}{j}
\renewcommand{\k}{k}

\subsection{Lab-frame calculation}

\subsubsection{Zero-order term}

The Zeeman shifts contribute only trivially to the frequency shift at zero order. The only difference comes from $\ehs{\ea \ga}{\i}_1$ which acquires an extra term $-i(\Delta_{\ea}-\Delta_{\ga})\ehs{\ea \ga}{\i}_0$.
%\Eq{Eq:seg_1_mag}{\ehs{\ea \ga}{\i}_1\propto & -i(\Delta_{\ea}-\Delta_{\ga})\ehs{\ea \ga}{\i}_0,}
This translates into an extra contribution to the zero-order shift of the form \Eq{Eq:zero_order_shift_mag}{\delta_0^{\g \e}= -\frac{\cos\theta}{2\pi N}\Bigg((\Delta_{\ea}-\Delta_{\ga})+\sum\limits_{i,j\neq i}U^{ji}_{\g\e}\Bigg).} In other words, the dipolar shift is simply shifted by the energy difference between $\g$ and $\e$ due to the Zeeman splitting, as expected.

\subsubsection{First-order term}

There are two independent contributions to the first-order term of the shift coming from the Zeeman splitting. One contribution comes from the product of two first-order coherences
\Eq{Eq:1st_term_first_order}{\frac{\ehS{x}_1\ehS{y}_1}{\ehS{x}_0^2} = \frac{\ehS{x}_1\ehS{y}_1}{\ehS{x}_0^2}\Bigg\rvert_{B=0} + (\Delta_{\ea}-\Delta_{\ga})\Bigg(\frac{\Gamma}{2}+\frac{\cos\theta}{N}\sum\limits_{i,j\neq i}E^{ji}_{\g\e}\Bigg).}
Here, the subscript $(\cdot)|_{B=0}$ indicates the expression $(\cdot)$ evaluated for zero magnetic field.
The other contribution comes from the second order coherence
\Eq{Eq:seg_2_mag}{\ehs{\ea \ga}{\i}_2=\ehs{\ea \ga}{\i}_2\big\rvert_{B=0} + 2i(\Delta_{\ea}-\Delta_{\ga})\Bigg(\Gamma\ehs{\ea \ga}{\i}_0-\Big(\ehs{\ea\ea}{\i}_0-\ehs{\ga\ga}{\i}_0\Big)\sum_{j\neq i}\G^{\j\i}_{\ma\na,\ma\na}\ehs{\ea \ga}{\j}_0\Bigg).}
Together, this yields
\Eq{Eq:2nd_term_first_order}{\frac{\ehS{y}_2}{\ehS{x}_0} - \frac{\ehS{y}_2}{\ehS{x}_0}\Bigg\rvert_{B=0} = 2\left(\frac{\ehS{x}_1\ehS{y}_1}{\ehS{x}_0^2}-\frac{\ehS{x}_1\ehS{y}_1}{\ehS{x}_0^2}\Bigg\rvert_{B=0}\right) \quad \implies\quad \delta^{\ga \ea}_1 = \delta^{\ga \ea}_1|_{B=0}.}
Thus, the magnetic field does not contribute to the dipolar shift at first order in the short-time expansion. However, note that the first-order approximation to the shift quickly becomes invalid as we increase the size of the Zeeman shifts. In the limit of large magnetic field it is better to switch from the lab to the rotating frame, as we show in the next section.

\subsection{Large Zeeman shift limit}

To take the strong magnetic field limit (i.e., $\mu_0|\mathbf{B}|\gg \Gamma$) it is convenient to rewrite the above equations of motion in the rotating frame of the field. Specifically, we introduce rotated variables as $\ehsr{a_m b_n}{i}= e^{i(\Delta_{a_m}-\Delta_{b_n})t}\ehs{a_m b_n}{i}$ with $a,b\in\{g,e\}$. Let us assume a strong magnetic field which generates large Zeeman shifts, such that $(\Delta_{e_\n}-\Delta_{\ea})\gg \Gamma$ and $(\Delta_{g_\m}-\Delta_{\ga})\gg \Gamma$ $\forall \, \m \neq \ma$ and $\n\neq\na$. Under the rotating-wave approximation, we disregard any fast oscillating terms appearing in the equations of motion of the rotated variables, i.e.,~$\eDt{e_\n}{e_\nn}\approx \delta_{\n,\nn}$ and $\eDt{g_\m}{g_\mm}\approx \delta_{\m,\mm}$ (where here $\delta$ stands for the Kronecker delta). Note that this type of oscillating phases usually appear multiplying each other. However, we can use this approximation because we are assuming that the ground and excited manifolds have different $g$-factors (i.e.,~$\delta_e\neq\delta_g$), such that $\eDt{e_\n}{e_\nn}$ and $\eDt{g_\m}{g_\mm}$ oscillate at vastly different frequencies.

The strong-field suppression of off-resonant processes in these equations implies that, compared to Eqs.~(\ref{Eq:seg_mag}-\ref{Eq:seg_sgg_mag}), fewer processes involving levels different from $\g$ and $\e$ contribute. As a consequence, the first-order coefficient of the dipolar shift becomes
\Eq{Eq:first_order_shift_strong_mag}{\bar{\delta}^{\g \e}_1 = \lim_{\mu_0|\mathbf{B}|\gg \Gamma}\delta^{\g \e}_1 = -\frac{1}{2\pi N}\sum_{i,j\neq i}\Bigg\{U^{ji}_{\g\e}\widetilde{\Gamma}_{\g\e}\left(\theta\right)+\Bigg(\sum_{k\neq i,j}\bar{W}^{kji}\left(\theta\right)+\bar{Q}^{ji}\left(\theta\right)\Bigg)\Bigg\},}
with the following definitions:
\Eq{bar_W}{\bar{W}^{kji}\left(\theta\right)\equiv W^{kji}_{p=(\alpha-\beta)}\left(\theta\right),}
\Eq{bar_Q}{\bar{Q}^{ji}\left(\theta\right)\equiv \frac{1}{2}\left(\sin^2(\theta/2)\sum_{p}A^{- ji;ij}_{g_{\beta+p}\e,g_{\beta+p} \e}-\cos^2(\theta/2)A^{- ji;ij}_{\g\e,\g\e}\right).}
In this limit, the three-body processes $\bar{W}^{kji}(\theta)$ involve only the transition interrogated (no sums over polarizations). The beyond-MF terms $\bar{Q}^{ji}(\theta)$ do maintain a summation over polarizations $p\neq (\beta-\alpha)$ and hence connect to other levels. Note, however, that all processes involve the excited state $\ea$ and no other excited state.

The above expressions show that to first order in time the dipolar shift of a multilevel system under a strong magnetic field still deviates from a naive two-level model (in which quantities are simply rescaled by the appropriate power of the CGC). The main difference is the first term in $\bar{Q}^{ji}(\theta)$  [Eq.~(\ref{bar_Q})], which describes both coherent and incoherent processes that involve $\ea$ and all ground states accessible through a single-photon transition of polarization $p$. 

%##################################################################%	
%------------ Checking adjacent-levels approximations   -----------%
%##################################################################%
\section{The role of non-adjacent levels}

In the main text we have pointed out that, due to the large number of equations to solve, only levels adjacent to the selected $\pi$-transitions were considered in the dynamics. We discuss here the validity of this approximation.

\begin{figure}[!h]
	\includegraphics[width=\textwidth]{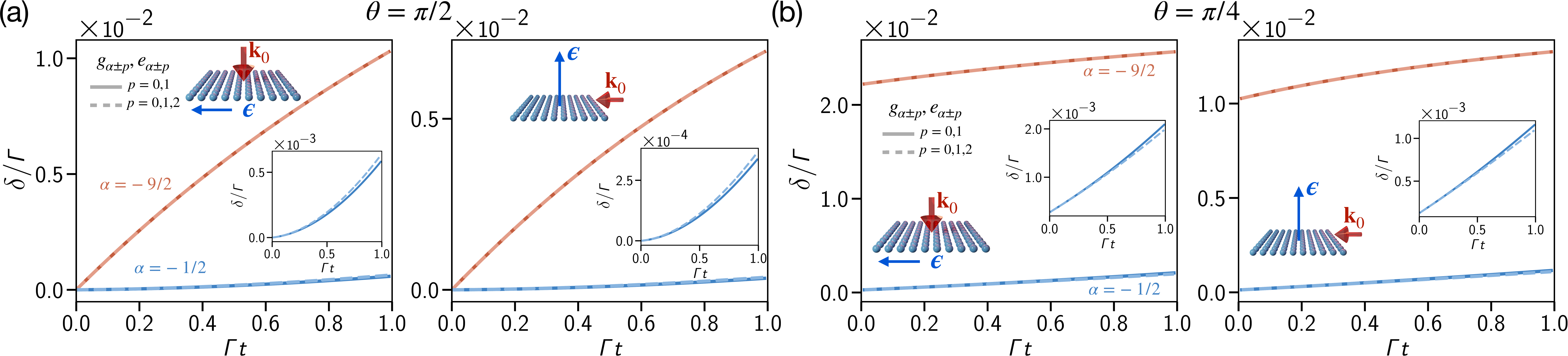} 
	\caption{Dipolar shift $\delta$ for different approximations of the internal level structure, with dynamics solved using cumulant simulations [approximation iii) in the main text]. We consider a system size of $N=3^2$ and the cartoons in each subplot show the direction of the interrogating laser and its polarization. We show simulations for a Ramsey pulse that interrogates a $\pi$-transition, $g_\alpha \rightarrow e_\alpha$, with pulse area (a) $\theta=\pi/2$, and (b) $\theta=\pi/4$. The simulations are performed for two different sets of levels: including only adjacent levels, $\{g_{\alpha\pm p},e_{\alpha\pm p}: p=0,1\}$ (full lines) and, for comparison, also including additional beyond-adjacent levels, $\{g_{\alpha\pm p},e_{\alpha\pm p}: p=0,1,2\}$ (dashed lines). Note that we are considering a system with angular momenta $F_e=F_g=9/2$, such that the levels included have to fulfill $|\alpha\pm p|\leq 9/2$. The curves are shown for $\alpha=-1/2$ and $-9/2$ (blue and red lines, respectively), where the insets provide a more detailed look at the former case.} \label{Fig:check_beyond_adjacent}
\end{figure}

For a given transition $g_\alpha \leftrightarrow e_\alpha$ the adjacent levels would be $g_{\alpha\pm1}$ and $e_{\alpha\pm1}$. All other levels, $g_{\alpha\pm p}$ and $e_{\alpha\pm p}$ with $p\neq 0,1$, will be called `non-adjacent'.
Figure~\ref{Fig:check_beyond_adjacent} presents a comparison of the dipolar shift obtained in cumulant simulations with ($p=0,1,2$: dashed lines) and without ($p=0,1$: solid lines) non-adjacent levels for a 2D system of $N=3^2$ atoms. Results for the $-9/2$ transition are shown in red, and results for $-1/2$ in blue.

The results of Fig.~\ref{Fig:check_beyond_adjacent} show that neglecting non-adjacent levels has no significant effect on the frequency shift over the dark times considered ($\Gamma t< 1$). As discussed in the main text, addressing the $-9/2$ transition leads to a time evolution practically identical to a two-level system. This is consistent with the insensitivity of the shift to non-adjacent levels shown in the plots. In contrast, we have shown in the main text that the $-1/2$ transition clearly deviates from a two-level system. However, we find that even in this case taking non-adjacent levels into account leads to barely observable differences in the shift, even at longer times. The difference is most clearly seen in the insets of Fig.~\ref{Fig:check_beyond_adjacent} which show the $-1/2$ case zoomed in.

To understand these findings, note that the zero and first-order expressions of the short-time expansion [see Eqs.~(\ref{Eq:zero_order_shift_SM}) and (\ref{Eq:first_order_shift_SM})] are fully independent from non-adjacent levels. Hence, non-adjacent levels are only expected to play a role at second or higher order in time. Furthermore, note that we are considering here examples without magnetic field. In the limit of a large magnetic field the non-adjacent levels fully decouple from the dynamics.
%This justifies analytically that indeed these levels play no role in the linear behavior of the dipolar shift obtained via a Ramsey spectroscopy protocol.
%\newpage

%##################################################################%	
%------ Local shifts: two-level and large-B limit comparison  -------%
%##################################################################%
\section{Local shifts comparison: two-level systems and large-B limit}

\begin{figure*}[!h]
	\includegraphics[width=\textwidth]{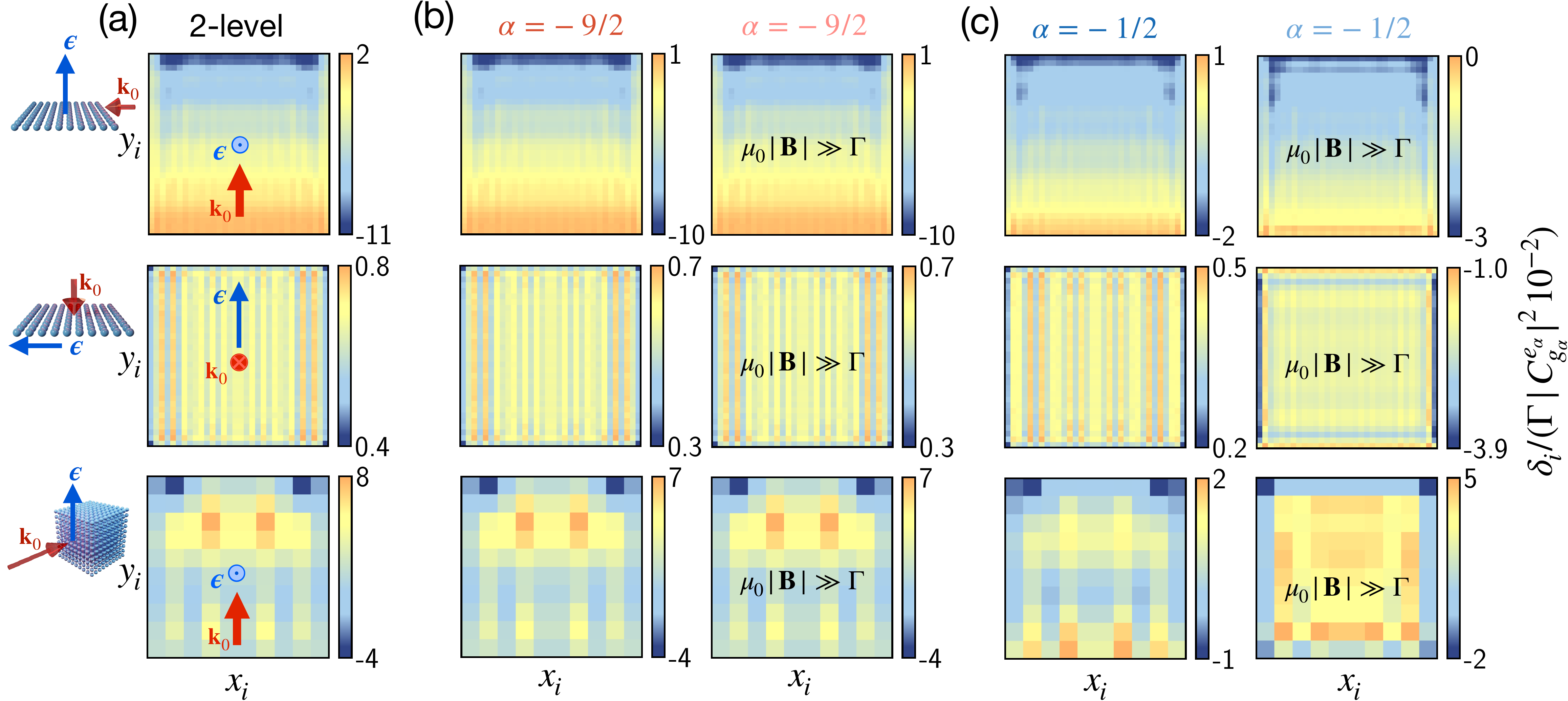} 
	\caption{Extended version of Fig.~\ref{Fig:local_shifts} from the main text (see corresponding caption). The additional columns offer the comparison with a two-level system, i.e., $|C^{e_\alpha}_{g_\alpha}|=1$, and with the large magnetic field limit for transition $\alpha=-9/2$ (first and third columns, respectively). The dipolar shift patterns of the first three columns are practically indistinguishable, evidencing the two-level nature of the $-9/2$ transition.} \label{Fig:extended_local_shifts_pi_2}
	\end{figure*}

We show in Fig.~\ref{Fig:extended_local_shifts_pi_2} the local shifts obtained for a more complete set of cases than in Fig.~\ref{Fig:local_shifts} in the main text. In particular, we added results for a two-level system (first column), for $\alpha=-9/2$ with a large magnetic field (third column), as well as for 2D geometries driven by a laser with $\mathbf{k}_0$ parallel (top row) and orthogonal (second row) to the atomic plane. The 3D plots that were shown in the main text are displayed here along with the new ones for ease of comparison.

The most important takeaway from these plots is the close similarity of the first three columns. This substantiates the claim made in the main text that the $-9/2$ transition can be well described as a two-level system.

It is worth pointing out that, for a $\pi/2$ pulse, as we consider here, the definition for the local shift provided in the main text, $\delta^{\g\e}_i \equiv \frac{1}{2\pi t} \arctan(\Xpct{\hat s^{y}_i}/\Xpct{\hat s^{x}_i})$, yields the same short-time expression as the one given in Eq.~(\ref{Eq:first_order_shift_SM}) but without the sum over $i$. In other words, for each atom $i$, $\delta^{\g\e}_i$ is obtained by simply evaluating the sums over $j\neq i$ in Eq.~(\ref{Eq:first_order_shift_SM}). This is not generally true for pulses with $\theta\neq \pi/2$.

%\newpage
%##################################################################%	
%------ First-order expression for Ramsey fringes contrast  -------%
%##################################################################%

\section{First-order expression for Ramsey fringes contrast}

The Ramsey fringes contrast was defined in the main text as \Eq{Eq:contrast}{\mathcal{C}^{\g \e}(t)\equiv \frac{1}{N}\sqrt{\ehS{x}^2(t)+\ehS{y}^2(t)}.} Following the same expansion in $\Gamma t$ as performed for the calculation of the frequency shift, the contrast up to first order in time is given by
\Eq{Eq:first_order_contrast}{\mathcal{C}^{\g \e}(\theta,t)\approx \sin\theta\Bigg[1-\Bigg(&\frac{\Gamma}{2}+\frac{\cos\theta}{N}\sum_{i,j\neq i}E^{ji}_{\g\e}\Bigg)t\Bigg].}
According to this expression, for a $\pi/2$ pulse and to first order in time, the fringes contrast decays solely through single-particle spontaneous emission. Collective effects for these short times are only relevant for Ramsey pulses with $\theta\neq \pi/2$. In these cases, the decay rate is modified by the collective term $E$, which accounts for phenomena such as subradiant or superradiant emission. Similarly to the classical-dipole energy $U$, the $E^{ji}_{\g\e}$ terms [see~Eq.~(\ref{Eq:U_and_E_def})] are also proportional to the CGC of the selected transition $\g\leftrightarrow\e$, and can thus be highly suppressed when $|C_{\g}^{\e}|^2\ll 1$.

\begin{figure}[!h]
	\includegraphics[width=\textwidth]{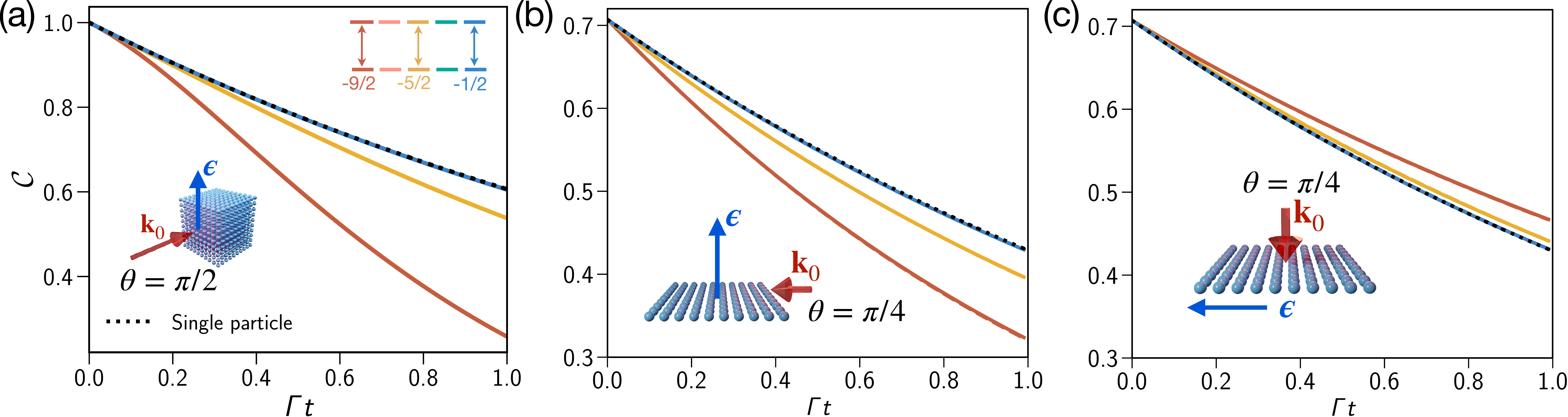} 
	\caption{Contrast decay of the Ramsey fringes. (a) 3D lattice with $N=10^3$ atoms and a $\pi/2$ pulse (see related FIG.~1 of the  main text). Notice that, for very short times and a $\pi/2$ pulse, the contrast decays independently of the transition selected, following the single-particle curve (dotted line), as described in Eq.~(\ref{Eq:first_order_contrast}). (b) and (c) 2D lattices with $N=8^2$ and $\pi/4$ pulse. Different wave vector and polarization directions lead to different sub- or superradiant decays of the contrast. In either case, the modification of the decay rate due to collective effects is strongest for the $-9/2$ transition.} \label{Fig:contrast}
\end{figure}

This dependence on the CGC is illustrated in Fig.~\ref{Fig:contrast}, which shows the contrast decay for the $-9/2$, $-5/2$, and $-1/2$ transitions and for three different geometrical configurations. These plots clearly show that the subradiant and superradiant modifications of the decay rate are most pronounced for the $-9/2$ transition, which has the largest CGC, whereas the $-1/2$ transition, which has the smallest CGC, is almost indistinguishable from the single-particle decay line. The $-5/2$ transition lies inbetween both extremes. Note that whether the decay is subradiant or superradiant depends on the overall sign of the sum over $E^{ji}_{\g\e}$ in Eq.~(\ref{Eq:first_order_contrast}). However, this sign is hard to predict from first principles, because the sum generally contains both positive and negative terms of different magnitudes which sensitively depend on details of the geometry.

%##################################################################%	
%------------------------ N-scaling --------------------------%
%##################################################################%

\section{Frequency shift scaling with system size}

In this section, we present results for the scaling of the dipolar shift with the total atom number (or system size) $N$, including more cases than shown in the main text. We provide results for the $-1/2$ transition in Figs.~\ref{Fig:first_order_prefactors_pi_4_center} and \ref{Fig:first_order_prefactors_pi_2_center}, and for the $-9/2$ transition in Figs.~\ref{Fig:first_order_prefactors_pi_4_left} and \ref{Fig:first_order_prefactors_pi_2_left}. Figures~\ref{Fig:first_order_prefactors_pi_4_center} and \ref{Fig:first_order_prefactors_pi_4_left} show the $\theta=\pi/4$ case and Figs.~\ref{Fig:first_order_prefactors_pi_2_center} and \ref{Fig:first_order_prefactors_pi_2_left} show the $\theta=\pi/2$ case. Each figure contains three columns, each of which corresponds to a different geometry configuration as indicated inside the top panels. In each figure, the top row shows in blue the total shift [computed with Eqs.~(\ref{Eq:zero_order_shift}) and (\ref{Eq:first_order_shift}) in the main text] and in black the beyond MF contribution to the shift [Eq.~(\ref{Eq:bmf}) in the main text]. The bottom row shows the individual contributions of the different terms in Eq.~(\ref{Eq:first_order_shift}). In all cases, solid lines indicate simulations done without magnetic field, whereas dashed, lighter lines correspond to the limit of a large magnetic field (the $U\widetilde{\Gamma}$ term is insensitive to magnetic fields, so dashed lines for this quantity are omitted). Note that in the bottom rows of Fig.~\ref{Fig:first_order_prefactors_pi_2_center}(a) and (c) the lines corresponding to the $W$ terms (yellow) cannot be seen, because their values are approximately zero and hence they are hidden behind other lines.

The main takeaways from these figures are the following. First, all figures show that the 2D array configuration with $\mathbf{k}_0$ parallel to the plane (center column) is the only case where the total shift appears to keep growing with $N$ for the range of values of $N$ considered. All other cases (left and right columns) seem to saturate. Second, for the $-1/2$ transition (Figs.~\ref{Fig:first_order_prefactors_pi_4_center} and \ref{Fig:first_order_prefactors_pi_2_center}) the beyond-MF term $\delta^{\g\e}_{\text{BMF}}$ (black lines in top row) accounts for a large share of the total shift in many of the cases presented. This shows the relevance of beyond-MF processes in this genuine multilevel case. Third, for $\pi/4$ pulses and the dark time considered here for all the plots ($\Gamma t=0.3$), the zero-order term is the leading contribution to the total shift. Lastly, for the $-9/2$ transition (Figs.~\ref{Fig:first_order_prefactors_pi_4_left} and \ref{Fig:first_order_prefactors_pi_2_left}) all dashed lines lie on top of the solid lines of the same color. This remarkable magnetic field insensitivity further substantiates the two-level character of the dynamics taking place when interrogating this transition.

The fact that the global shift only appears to grow with $N$ for the 2D configuration with $\mathbf{k}_0$ parallel to the atomic plane can be qualitatively understood from geometrical considerations. The key element is that the polarization $\boldsymbol\epsilon$, i.e.~the quantization axis, is perpendicular to the atomic plane. Because of this, the dipolar interaction coefficients involving $\pi$-transitions become isotropic within the 2D plane (which is not the case when $\boldsymbol\epsilon$ is parallel to the atomic plane). In the far-field limit ($k_0r_{ij}\gg1$) they are given by
\begin{align}
    \Delta^{ij}_{g_\alpha e_\alpha, g_\alpha e_\alpha} = \frac{3\Gamma}{4} \left( C^{e_\alpha}_{g_\alpha} \right)^2 \mathrm{Re} \frac{e^{ik_0r_{ij}}}{k_0r_{ij}},\qquad\qquad
    \Gamma^{ij}_{g_\alpha e_\alpha, g_\alpha e_\alpha} = \frac{3\Gamma}{4} \left( C^{e_\alpha}_{g_\alpha} \right)^2 \mathrm{Im} \frac{e^{ik_0r_{ij}}}{k_0r_{ij}}.
\label{eq:DGamma_2D}
\end{align}
In the short-time expansion presented above, these interaction coefficients appear in convoluted sums over the whole 2D array. The asymptotic behavior of these sums can be studied by approximating them through integrals. Usually, a 3D integral over the dipolar interactions vanishes due to the anisotropy of the dipolar interactions. This explains the saturation of the shift observed for the 3D case in Figs.~\ref{Fig:first_order_prefactors_pi_4_center}, \ref{Fig:first_order_prefactors_pi_2_center}, \ref{Fig:first_order_prefactors_pi_4_left}, and \ref{Fig:first_order_prefactors_pi_2_left}. However, since the interaction coefficients in Eq.~(\ref{eq:DGamma_2D}) are isotropic within the 2D atomic plane, the 2D integrals can show nontrivial scaling with $N$ due to the long-range tail $1/(k_0r_{ij})$ of the dipolar interactions. Note, however, that the system sizes employed in the numerical results of Figs.~\ref{Fig:first_order_prefactors_pi_4_center}, \ref{Fig:first_order_prefactors_pi_2_center}, \ref{Fig:first_order_prefactors_pi_4_left}, and \ref{Fig:first_order_prefactors_pi_2_left} are not sufficiently large to observe a clean scaling with $N$.

As an example, we can consider the classical energy term $\mathcal{U}_\alpha \equiv \frac{1}{N}\sum_{i,j\neq i} U^{ji}_{g_\alpha e_\alpha}$, which appears in both the zero and the first-order shift expressions, see Eqs.~(\ref{Eq:zero_order_shift_SM}) and (\ref{Eq:first_order_shift_SM}). For the 2D configuration with $\mathbf{k}_0$ parallel to the atomic plane we have, in the $k_0r_{ij}\gg1$ limit, $U^{ji}_{g_\alpha e_\alpha} \propto \mathrm{Re}[ e^{ik_0r_{ij}-i\mathbf{k}_0\cdot \mathbf{r}_{ij}}/(k_0r_{ij})]$.
In the large-$N$ limit we can assume that $\sum_{j\neq i} U^{ji}_{g_\alpha e_\alpha}$ is approximately independent of $i$, and substitute the remaining sum by a 2D integral as (up to constants) $\mathcal{U}_\alpha \sim \mathrm{Re} \int_0^{2\pi} d\varphi \int_\varepsilon^L dr e^{ir(1-\cos\varphi)}$, where $\varepsilon>0$ is an ultraviolet cutoff and $L^2\sim N$. A simple analysis then predicts $\mathcal{U}\sim N^{1/4}$. An analogous calculation can be performed for $\mathcal{E}_\alpha \equiv \frac{1}{N}\sum_{i,j\neq i} E^{ji}_{g_\alpha e_\alpha}$, which also yields $\mathcal{E}\sim N^{1/4}$.

\newcommand{\figsize}{0.85}

\begin{figure}[!h]
	\includegraphics[width=\figsize\textwidth]{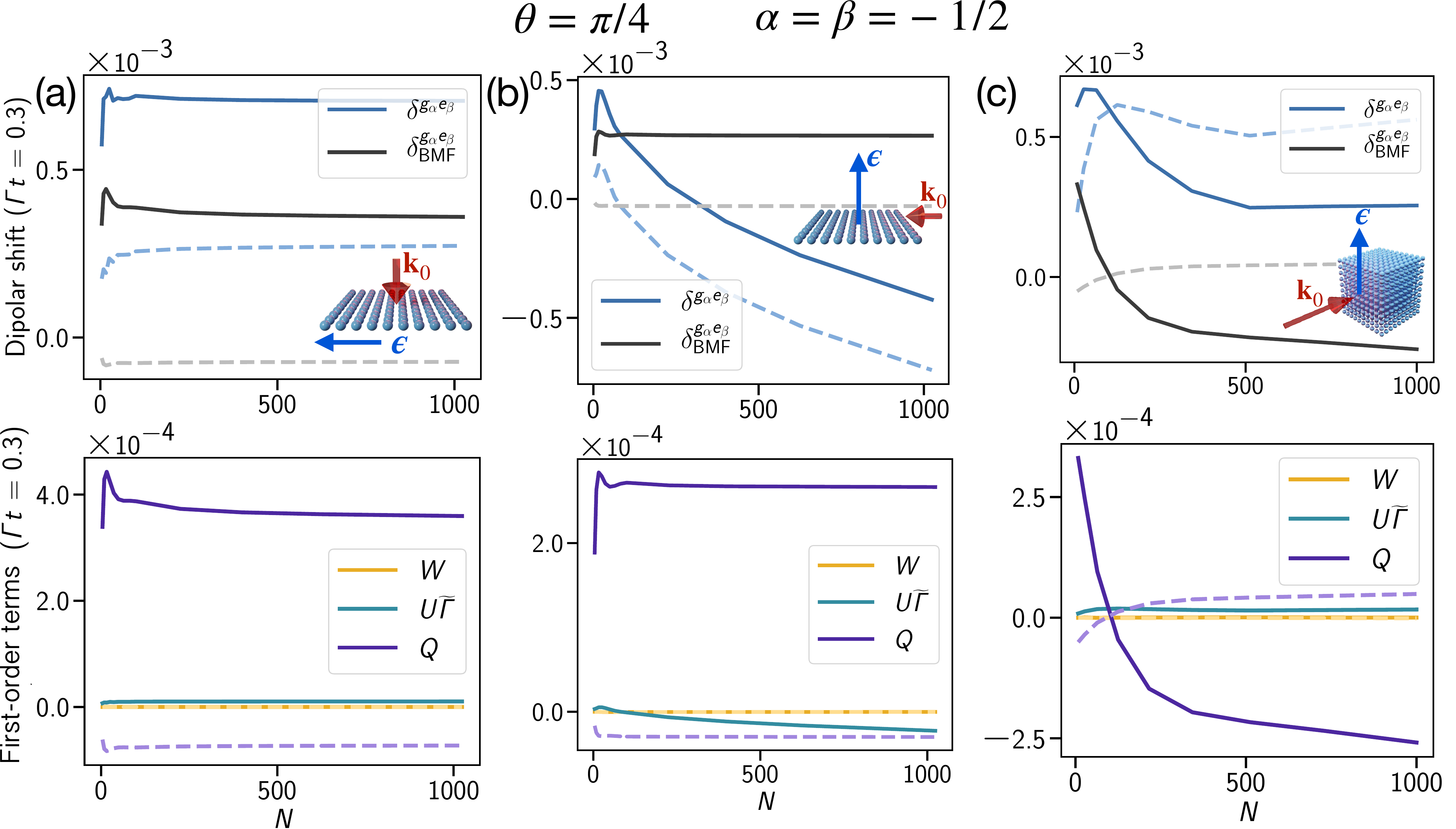} 
	\caption{Scaling with system size $N$ at time $\Gamma t=0.3$ of the total shift (blue lines, top), beyond-MF term (black lines, top), and first-order terms of Eq.~(\ref{Eq:first_order_shift}) (bottom). The geometry considered for each column of plots is depicted in the top plots, along with the associated interrogation laser direction and polarization. The dashed, lighter lines correspond to the limit of a strong magnetic field, whereas solid lines are computed for zero magnetic field. In this figure, we considered addressing the $\pi$-polarized $-1/2$ transition with a pulse area of $\theta=\pi/4$. Without magnetic field, the total shift for the times considered is dominated for all geometries by a combination of the zero-order shift and the beyond-mean-field, first-order term $Q$. All other first-order terms appear negligible. With a strong magnetic field, the $Q$ term is strongly modified, evidencing the multilevel nature of the system.} \label{Fig:first_order_prefactors_pi_4_center}
\end{figure}

\begin{figure}[!h]
	\includegraphics[width=\figsize\textwidth]{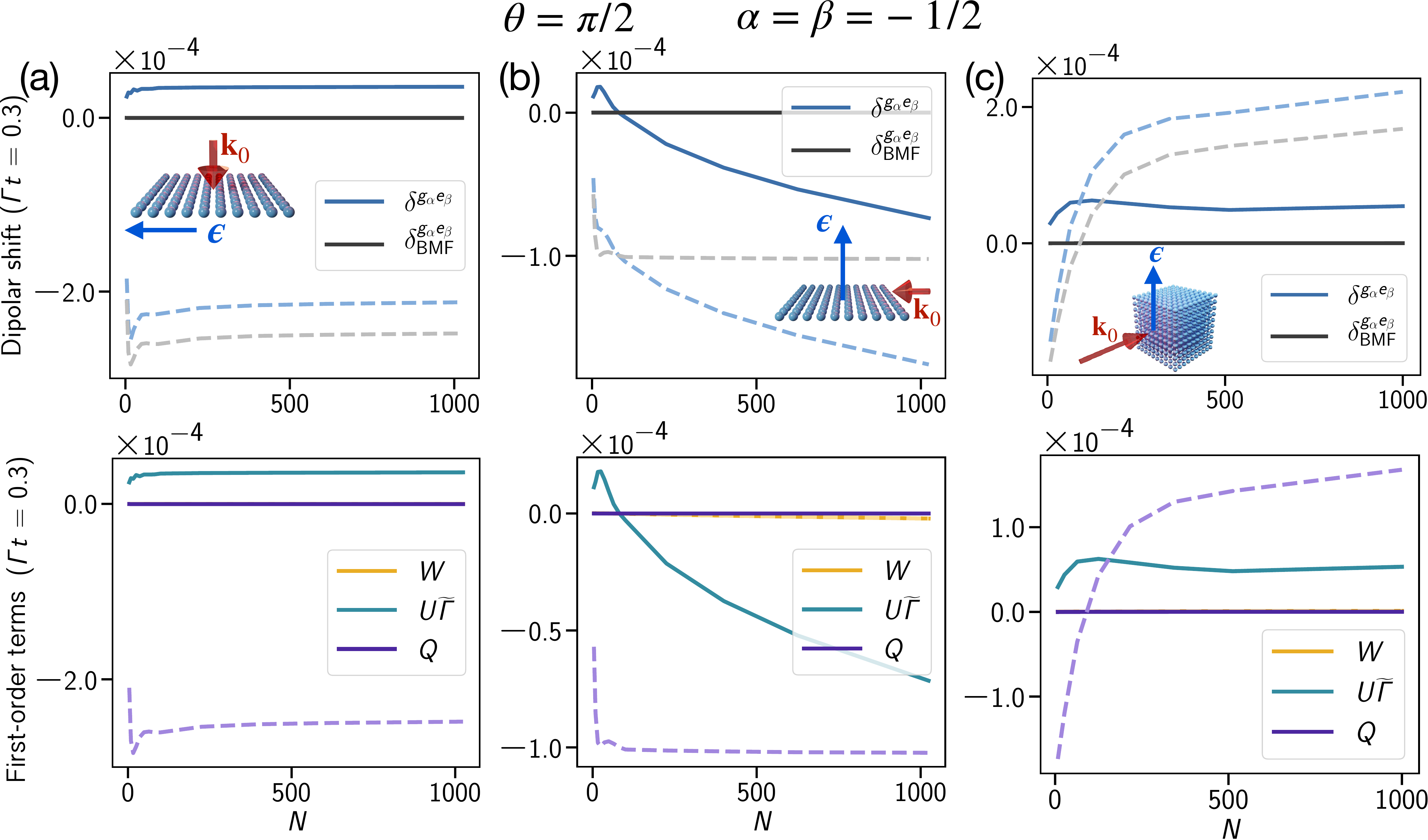} 
	\caption{Scaling with system size $N$ at time $\Gamma t=0.3$ of the total shift (blue lines, top), beyond-MF term (black lines, top), and first-order terms of Eq.~(\ref{Eq:first_order_shift}) (bottom). The geometry considered for each column of plots is depicted in the top plots, along with the associated interrogation laser direction and polarization. The dashed, lighter lines correspond to the limit of a strong magnetic field, whereas solid lines are computed for zero magnetic field. In this figure, we considered addressing the $\pi$-polarized $-1/2$ transition with a pulse area of $\theta=\pi/2$. The leading term in the absence of a magnetic field is $\widetilde{\Gamma}$ times the total classical dipole energy $U$ [first line in Eq.~(\ref{Eq:first_order_shift})]. The application of a strong magnetic field, on the other hand, makes the beyond-mean-field, two-body correlation term $Q$ the dominant one.} \label{Fig:first_order_prefactors_pi_2_center}
	\end{figure}

\begin{figure}[!h]
	\includegraphics[width=\figsize\textwidth]{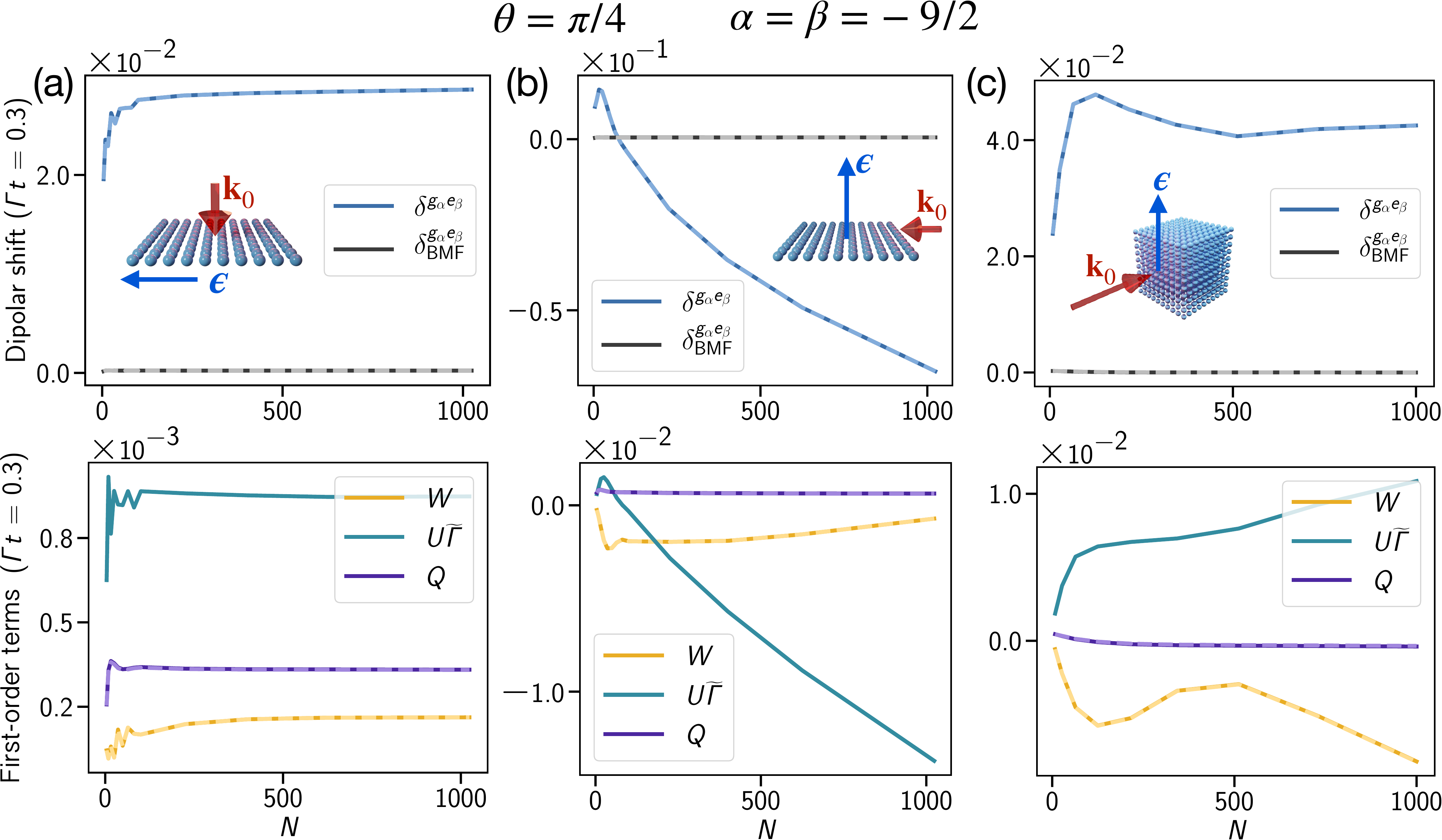} 
	\caption{Scaling with system size $N$ at time $\Gamma t=0.3$ of the total shift (blue lines, top), beyond-MF term (black lines, top), and first-order terms of Eq.~(\ref{Eq:first_order_shift}) (bottom). The geometry considered for each column of plots is depicted in the top plots, along with the associated interrogation laser direction and polarization. The dashed, lighter lines correspond to the limit of a strong magnetic field, whereas solid lines are computed for zero magnetic field. In this figure, we considered addressing the $\pi$-polarized $-9/2$ transition with a pulse area of $\theta=\pi/4$. For the dark time considered ($\Gamma t=0.3$), the shift is mostly dominated by the zero-order contribution, similarly to the $-1/2$ case with $\pi/4$. Here, however, the first-order terms are mostly dominated by $U\widetilde{\Gamma}$ and the three-body contributions $W$, since they are not strongly suppressed as in the $-1/2$ case.  All dashed curves (with a strong bias magnetic field) lie on top of the corresponding full-line curves (no magnetic field), confirming the two-level nature of the $-9/2$ transition.} \label{Fig:first_order_prefactors_pi_4_left}
	\end{figure}

\begin{figure}[!h]
	\includegraphics[width=\figsize\textwidth]{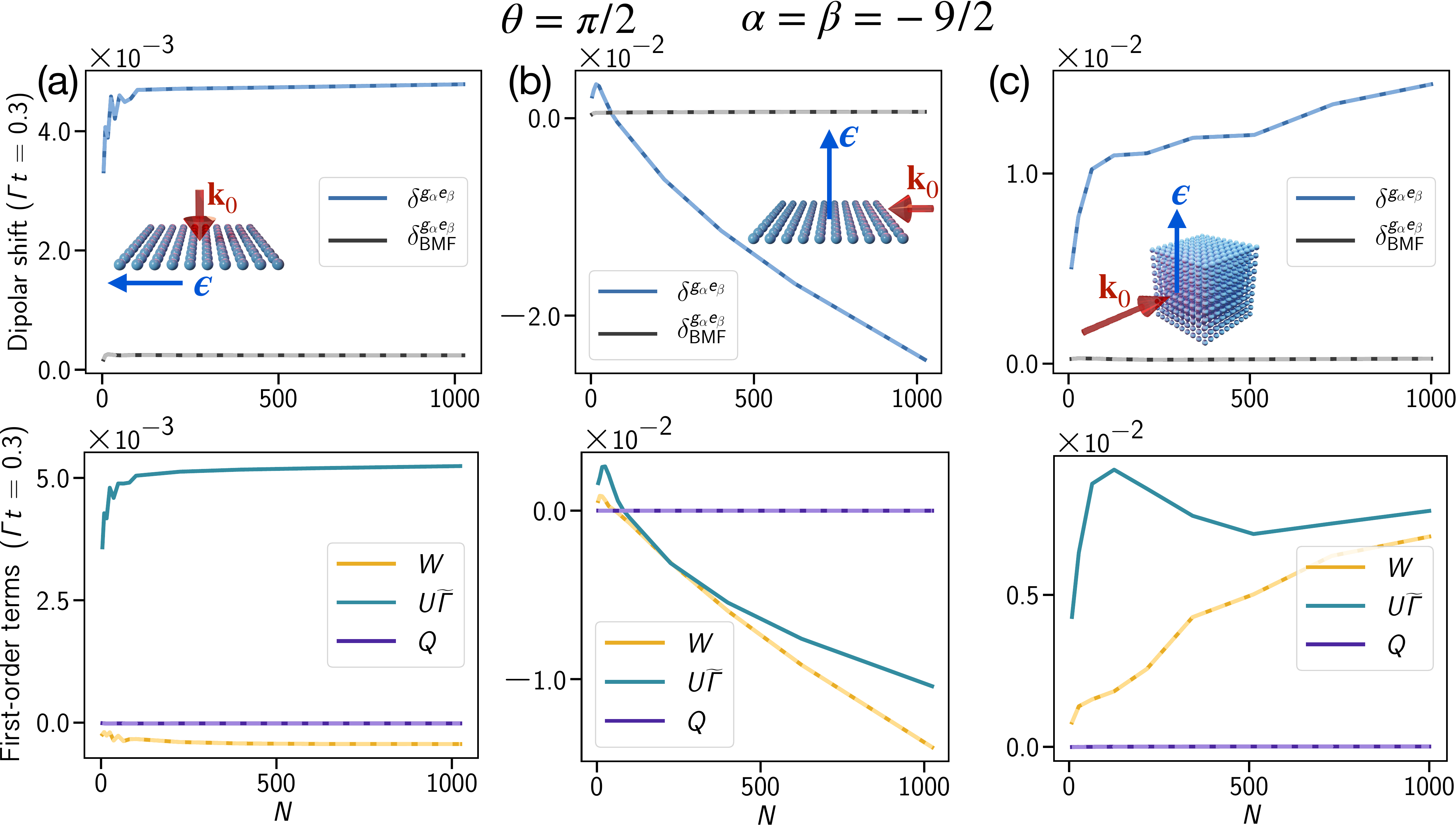} 
	\caption{Scaling with system size $N$ at time $\Gamma t=0.3$ of the total shift (blue lines, top), beyond-MF term (black lines, top), and first-order terms of Eq.~(\ref{Eq:first_order_shift}) (bottom). The geometry considered for each column of plots is depicted in the top plots, along with the associated interrogation laser direction and polarization. The dashed, lighter lines correspond to the limit of a strong magnetic field, whereas solid lines are computed for zero magnetic field. In this figure, we considered addressing the $\pi$-polarized $-9/2$ transition with a pulse area of $\theta=\pi/2$. This case is dominated by $U\widetilde{\Gamma}$ [first line in Eq.~(\ref{Eq:first_order_shift})] and the typically comparable three-body contributions $W$ [except for the two-dimensional geometry considered in (a)]. In (b), for $N=8^2$, the total shift ($\approx 1.5\times 10^{-3}$) is comparable to $\delta_\text{BMF}$, which justifies the large deviation between MF and cumulant simulations in Fig.~\ref{Fig:shift_BMF_scaling}(c) of the main text. All dashed curves (with a strong bias magnetic field) lie on top of the corresponding full-line curves (no magnetic field), confirming the two-level nature of the $-9/2$ transition.} \label{Fig:first_order_prefactors_pi_2_left}
	\end{figure}

\end{document}